\documentclass[12pt]{article}
\usepackage{amssymb,amsmath,cite,mathdots}

\def\beq{\begin{equation}}
\def\eeq{\end{equation}}
\def\bea{\begin{eqnarray*}}
\def\eea{\end{eqnarray*}}
\def\nn{\nonumber}

\def\ket#1{\left| #1\right\rangle}

\def\N0{ {\mathbb Z}_{+} }
\def\v0{ |d,r) }

\def\proof#1{{\bf Proof:} #1 $\blacksquare$ \medskip}

\newtheorem{lemma}{Lemma}

\newtheorem{thm}{Theorem}

\begin{document}

\begin{center}
{\large\bf Matrix elements for type 1 unitary irreducible representations of the Lie superalgebra $gl(m|n)$}\\
~~\\

{\large Mark D. Gould, Phillip S. Isaac and Jason L. Werry}\\
~~\\

School of Mathematics and Physics, The University of Queensland, St Lucia QLD 4072, Australia.
\end{center}

\begin{abstract}
Using our recent results on eigenvalues of invariants associated to the Lie superalgebra
$gl(m|n)$, we use characteristic identities to derive explicit matrix element formulae 
for all $gl(m|n)$ generators, particularly {\em non-elementary} generators, on finite
dimensional type 1 unitary irreducible representations. 
We compare our results with existing works that deal with only subsets of the class of type 1
unitary representations, all of which only present explicit matrix elements
for elementary generators. Our work therefore provides an important
extension to existing methods, and thus highlights the strength of our techniques which
exploit the characteristic identities.
\end{abstract}

%

\section{Introduction}

This is the second paper in a series aimed at deriving matrix elements of elementary and non-elementary 
generators of
finite dimensional unitary irreducible representations for the Lie superalgebra $gl(m|n)$. Such
representations were classified in the work of Gould and Zhang
\cite{ZhaGou1990,GouZha1990}, although the concept of a conjugation operation (necessary
to understand unitary representations) was well known before this thanks to the work of
Scheunert, Nahm and Rittenberg \cite{SNR1977}. There are two types of finite dimensional
irreducible unitary representations of $gl(m|n)$ that are defined depending on the
sesquilinear form that exists on the module (see Section \ref{class} below for more
details). In this paper, we consider only the irreducible type 1 unitary representations,
and make use of the classification results of \cite{ZhaGou1990,GouZha1990}. The matrix
elements of the irreducible unitary representations of type 2 will be dealt with in a
subsequent article.

The first of our papers in the series \cite{GIW1} was aimed at constructing invariants associated
with $gl(m|n)$, and determining analytic expressions for their eigenvalues, which paves
the way for the current article. Indeed, as we shall see, we
rely heavily on some of the results obtained in \cite{GIW1}. 
Another important motivation for the first article was to highlight the innovative
techniques involving characteristic identities
\cite{Green1971,BraGre1971,Green1975,OBCantCar1977,Gould1985}. We are certainly of the opinion that
these characteristic identities are a valuable yet underestimated (perhaps even unknown)
mathematical tool, and over the course of this series of papers we aim to convince readers
of their usefulness and importance. Characteristic identities associated to Lie
superalgebras have been studied in the work of Green and Jarvis
\cite{JarGre1979,GreJar1983} and Gould \cite{Gould1987}. 

Our previous paper \cite{GIW1} gives a reasonably detailed survey of the current literature on the
subject, not only concerning invariants, but also regarding the matrix element formulae
themselves. We therefore direct the reader to that article for the broad setting of the current
work. The types of matrix element formulae that we will present here were first
written down by Gelfand and Tsetlin \cite{GT1950,GT1950b} for the general linear and
orthogonal Lie algebras, and many works
followed. Our approach is of a similar nature to the work of Baird and Biedenharn
\cite{BB1963}, in the sense that we are interested in advocating the methods
used, as well as the matrix element formulae themselves, which are derived for all
(i.e. both elementary {\em and} non-elementary) generators. 
We find that the results presented
in this paper, particularly in Sections \ref{meg} and \ref{resultsum}, are of such generality that they
encompass the results of previous works of Palev \cite{Palev1987,Palev1989}, Stoilova and
Van der Jeugt \cite{StoiVan2010} and Molev \cite{Molev2011} (also Tolstoy et al.
\cite{TolIstSmi1986}) for the case of the elementary generators. 
Hence our methods appear to unite other approaches, and certainly
contexts. More significantly, however, the current article extends these works to derive
explicit expressions for the matrix elements of the non-elementary generators. This is
important not only for analytic completeness, but for computational efficiency. For
instance, the computational time required to compute matrix elements of non-elementary
generators via (anti-)commutators grows exponentially with the dimensions of the
representation and the Lie superalgebra. Thus in any concrete application, having explicit matrix element
formulae of the non-elementary generators is essential for maximising algorithmic
efficiency. This has indeed been demonstrated for prototype applications of normal Lie
algebras in the context of the nuclear shell model \cite{Moshinsky1968} and quantum
chemistry \cite{Paldus1974,Paldus1975,Paldus1976,Shavitt1977,Shavitt1978}.

We also give explicit statements of branching rules for the case of all type 1 unitary
representations corresponding to a basis symmetry adapted to the chain of subalgebras
$$
gl(m|n+1)\supset gl(m|n)\supset \cdots \supset gl(m|1)\supset gl(m)\supset
gl(m-1)\supset \cdots \supset gl(1).
$$
Having such explicit rules for the decomposition of an irreducible type 1 unitary
$gl(m|n+1)$ highest weight module into a direct sum of irreducible type 1 unitary
$gl(m|n)$ highest weight modules lends itself to representing the basis vectors using the
familiar combinatorial Gelfand-Tsetlin (GT) patterns generalised to the case of Lie
superalgebras. Such patterns have already appeared
throughout the literature for a variety of Lie superalgebras, particularly for $gl(m|n)$ in
the articles \cite{Palev1987,Palev1989,StoiVan2010,Molev2011,TolIstSmi1986}. The
branching rules presented in the current article simplify the conditions
satisfied by the weight labels in such GT patterns to encompass all type 1 unitary highest
weight modules of $gl(m|n)$.

The paper is organised as follows. Section \ref{rev} provides a brief review of the
context and important notations used throughout the paper. Section \ref{class} then
summarises the classification results of Gould and Zhang \cite{ZhaGou1990,GouZha1990}
pertaining to the unitary representations of $gl(m|n)$ in particular. Specifically for the
type 1 unitary representations, in Section \ref{repcat} we present three main subclasses of representations, which
correspond to certain cases already presented in the literature, and highlight a case
which is genuinely new, although somewhat simplistic. The details of the branching rules
are given in Section \ref{branch}, followed by a construction of the explicit matrix
element formulae derived using characteristic identities in Section \ref{meg}. The
explicit matrix element formulae are then presented in their entirety in Section
\ref{resultsum}. To help
solidify some of the concepts encountered by the reader, we present an example in Section
\ref{example} that deals with an arbitrary irreducible type 1 unitary representation of
$gl(2|2)$.


\section{Review} \label{rev}

In this article, we adopt the same graded index notation used in \cite{GIW1}, with Latin
indices $1\leq i,j,\ldots\leq m$ assumed to be even, and Greek indices
$1\leq\mu,\nu,\ldots\leq n$ taken to be odd. The parity of the index is characterised by
$$
(i)=0,\ \ (\mu)=1.
$$
As in \cite{GIW1}, where convenient we sometimes use indices $1\leq p,q,r,s\leq m+n$, in
which case if $p=1,\ldots,m$ then we write
$(p)=0$, and if $p=m+\mu$ for some $\mu=1,\ldots,n$ then we write $(p)=(\mu)=1.$ 
In particular, the $gl(m|n)$ generators $E_{pq}$ satisfy the graded commutation relations
$$
[E_{pq},E_{rs}] 
= \delta_{qr}E_{ps} - (-1)^{[(p)+(q)][(r)+(s)]}\delta_{ps}E_{rq}.
$$

We let $H$ denote the Cartan subalgebra of $gl(m|n)$, made up of the mutually commuting
generators $E_{pp}$. Using the same notation as \cite{GIW1},
we may express any weight $\Lambda\in H^*$ 
in terms of fundamental weights $\varepsilon_i$ and $\delta_\mu$ as introduced by Kac \cite{Kac1977}, 
as
$$
\Lambda = \sum_{i=1}^m \Lambda_i\varepsilon_i + \sum_{\mu=1}^n \Lambda_\mu\delta_\mu,
$$
where the coefficients $\Lambda_p$ are the eigenvalues of the generators $E_{pp}$ on a
corresponding weight vector.

Let $\psi$ be a homogeneous graded intertwining operator
$$
\psi:V(\varepsilon_1)\otimes V \rightarrow W,
$$
where $V$ and $W$ are $gl(m|n)$ modules, and $V(\varepsilon_1)$ is the module
corresponding to the fundamental vector representation with highest weight
$\varepsilon_1$. Setting $\{ e^r\ |\ 1\leq r\leq m+n\}$ as a homogeneous basis
for $V(\varepsilon_1)$, the set of components $\{ \psi^r\}$, collectively referred to as a vector
operator, has an action on $V$ given by
$$
\psi^r v = \psi(e^r\otimes v),\ \ \forall v\in V.
$$
In terms of the graded commutator, the transformation law for vector operators can be
expressed \cite{GIW1} as
\begin{align}
[E_{pq},\psi^r] = (-1)^{(\psi)[(p)+(q)]} \delta^r_{\ q}\psi^p.
\nn
\end{align}

Similarly, let $\phi$ be the graded intertwining operator
$$
\phi:V(\varepsilon_1)^*\otimes V \rightarrow W,
$$
where $V(\varepsilon)^*$ denotes the dual vector representation.
Setting $\{ e_r\ |\ 1\leq r\leq m+n\}$ as a homogeneous basis
for $V(\varepsilon_1)^*$, the set of components $\{ \phi_r\}$ define a contragredient
vector operator, which transforms under graded commutation according to
\begin{align}
[E_{pq},\phi_r] = -(-1)^{[(\phi)+(q)][(p)+(q)]} \delta_{pr}\phi_q.
\nn
\end{align} 

In this article, we are interested in {\em odd} vector and contragredient vector operators
(with $(\psi)=1=(\phi)$). 
In this case, if $\chi^r$ and $\psi^s$ are two odd vector operators, and $\phi_r$ and
$\xi_s$ are two odd contragredient vector operators, then the following are worth noting:
\begin{align}
[E_{pq},\chi^r\psi^s] &= [E_{pq},\chi^r]\psi^s +
(-1)^{[(p)+(q)][1+(r)]}\chi^r[E_{pq},\psi^s], 
\nn\\
[E_{pq},\phi_r\xi_s] &= [E_{pq},\phi_r]\xi_s +
(-1)^{[(p)+(q)][1+(r)]}\phi_r[E_{pq},\xi_s]. 
\nn
\end{align}

As in \cite{GIW1}, for $1\leq p,q\leq m+n$, we define a matrix $\bar{{\cal A}},$ associated to the
vector representation $\pi_{\varepsilon_1}$, with entries
\begin{align}
\bar{{\cal A}}_p^{\ q} = -(-1)^{(p)(q)}E_{qp}, \label{adjointA}
\end{align}
which can be shown \cite{GIW1} to satisfy the characteristic identity
\begin{align}
\prod_{i=1}^m (\bar{{\cal A}} - \bar{\alpha}_i) \prod^n_{\mu = 1} (\bar{{\cal A}} - \bar{\alpha}_\mu) =
0
\label{adjcharid}
\end{align}
when acting on an irreducible $gl(m|n)$ module $V(\Lambda)$.
The characteristic roots $\bar{\alpha}_i, \bar{\alpha}_\mu$ are given in terms of the
highest weight labels $\Lambda_r$ by
\begin{align}
\bar{\alpha}_i &= i - 1 -\Lambda_i,\ \ 1\leq i\leq m,\label{evenalphabar}\\
\bar{\alpha}_\mu &= \Lambda_\mu + m + 1 - \mu, \ \ 1\leq \mu\leq n.\label{oddalphabar}
\end{align}

We may also define a matrix ${\cal A},$ which is associated to the (triple) dual vector
representation $\pi_{\varepsilon_1}^{***},$ with entries
\begin{equation}
{\cal A}^p_{\ q} = (-1)^{(p)} E_{pq},
\label{charmat}
\end{equation}
which can also be shown to satisfy the characteristic equation
\begin{align}
\prod^m_{i=1} ({\cal A} - \alpha_i) \prod^n_{\mu=1} ({\cal A} - \alpha_\mu) = 0, 
\label{veccharid}
\end{align}
where the characteristic roots $\alpha_i,\alpha_\mu$ are given by
\begin{align}
\alpha_i  &= \Lambda_i + m - n - i,\ \ 1\leq i\leq m, \label{evenalpha}\\
\alpha_\mu  &= \mu-\Lambda_\mu - n,\ \ 1\leq \mu \leq n.\label{oddalpha}
\end{align}

Using equations (\ref{adjcharid}) and (\ref{veccharid}) respectively, we may construct
projection operators
\begin{align}
\bar{P}[r] &=  \prod^{m+n}_{k \neq r} \left( \frac{\bar{{\cal A}} - \bar{\alpha}_k}{\bar{\alpha}_r
- \bar{\alpha}_k} \right),\nn\\
P[r] &= \prod^{m+n}_{k \neq r} \left( \frac{{\cal A} - \alpha_k}{\alpha_r - \alpha_k}
\right),\nn
\end{align}
and then use these projections to obtain spectral resolutions of the vector and
contragredient vector operators respectively:
\begin{align}
\psi^p &= \sum^m_{i=1} \psi[i]^p + \sum^n_{\mu = 1} \psi[\mu]^p, \nn\\
\phi_p &= \sum^m_{i=1} \phi[i]_p + \sum^n_{\mu = 1} \phi [\mu]_p.\nn
\end{align}
In the above we have used the shift components $\psi[r]$ and $\phi[r]$, expressible in
terms of the appropriate projections (summation over repeated indices assumed) as
\begin{align}
\psi[r]^p &= \psi^q \bar{P}[r]_q^{\ p} = P[r]^p_{\
q}\psi^q,\nn\\
\phi[r]_p &= \bar{P}[r]_p^{\ q}\phi_q = (-1)^{(p)+(q)}\phi_q
P[r]^q_{\ p}.\nn
\end{align}
The shift components $\psi[r]$ and $\phi[r]$ effect the following shifts in the
representations labels $\Lambda$:
\begin{align}
\psi[r]: \Lambda_q \rightarrow \Lambda_q + \delta_{rq},\nn\\
\phi[r]: \Lambda_q \rightarrow \Lambda_q - \delta_{rq}.\nn
\end{align}


%
%

\section{Classification of unitary irreducible representations of $gl(m|n)$}
\label{class}

In this section we summarise the classification results of Gould and Zhang \cite{ZhaGou1990,GouZha1990} 
pertaining to finite dimensional irreducible unitary representations of $gl(m|n)$. 

On every irreducible, finite dimensional $gl(m|n)$-module $V(\Lambda)$, there exists a 
non-degenerate sesquilinear form $\langle\ |\ \rangle_\theta$ (unique up to a scalar
multiple) with the distinguished property
$$
\langle E_{pq}v | w\rangle_\theta = (-1)^{(\theta - 1)[(p)+(q)]}\langle v|E_{qp} w\rangle_\theta,
$$
with $\theta = 1$ or 2 relating to two inequivalent forms. The irreducible, finite
dimensional module $V(\Lambda)$ is said to be type $\theta$ unitary if $\langle\ |\
\rangle_\theta$ is positive definite on $V(\Lambda)$, and the corresponding
representation is said to be type $\theta$ unitary. Equivalently, for a finite dimensional unitary
irreducible representation $\pi$, we require an inner product such that the
linear operators $\pi(E_{pq})$ satisfy the hermiticity condition,
\begin{equation}
\left[ \pi(E_{pq}) \right]^\dagger = (-1)^{(\theta - 1)[(p)+(q)]} \pi(E_{qp}),
\label{hermit}
\end{equation}
where $\dagger$ denotes normal Hermitian conjugation defined by 
$$
\left( \left[ \pi(E_{pq}) \right]^\dagger \right)_{\alpha\beta} = {\left[
\overline{\pi(E_{pq})} \right]}_{\beta\alpha},
$$
with $\overline{A}$ denoting the matrix with complex entries conjugate to those of $A$.

Using the bilinear form on the fundamental weights
$$
(\varepsilon_i,\varepsilon_j) = \delta_{ij},\ \ (\varepsilon_i,\delta_\mu)=0, \ \ (\delta_\mu,\delta_\nu) =
-\delta_{\mu\nu},
$$
that was discussed in \cite{GIW1}, 
we have a non-degenerate bilinear form on the weights given by
\begin{equation}
\left( \Lambda,\Lambda'\right) = \sum_{i=1}^m\Lambda_i{\Lambda}_i' -
\sum_{\mu=1}^n\Lambda_\mu{\Lambda}_\mu'.
\nonumber
\end{equation}
We also have a distinguished system of simple roots given by 
\begin{equation}
\Phi = \left\{ \left. \varepsilon_i-\varepsilon_{i+1},\ \varepsilon_m-\delta_1,\
\delta_\mu-\delta_{\mu+1}\ \right| \ 1\leq i<m,\ 1\leq \mu<n \right\}.
\label{simpleroots}
\end{equation}
Furthermore, the sets of even and odd positive roots are given, respectively, by
\begin{align}
\Phi_0^+ &= \{ \varepsilon_i - \varepsilon_j\ |\ 1\leq i<j\leq m\} \cup \{\delta_\mu -
\delta_\nu\ |\ 1\leq \mu<\nu\leq n\},\nn\\
\Phi_1^+ &= \{ \varepsilon_i - \delta_\mu\ |\ 1\leq i\leq m,\ 1\leq\mu\leq n\}. \nn
\end{align}
Define $\rho$ to be the graded half-sum of positive roots, i.e.
\begin{align}
\rho &= \frac12 \sum_{\alpha\in\Phi_0^+}\alpha - \frac12\sum_{\beta\in\Phi_1^+}\beta\nn\\
&= \frac12 \sum_{j=1}^m(m-n-2j+1)\varepsilon_j +
\frac12\sum_{\nu=1}^n(m+n-2\nu+1)\delta_\nu.
\nonumber
\end{align}

Before proceeding, we remind the reader of some useful results. It is well known
\cite{SNR1977} that any
finite dimensional type 1 unitary representation is completely reducible, and that the
tensor product of two type 1 unitary representations is also type 1 unitary. We therefore
have \cite{SNR1977}:

\begin{thm}
\label{type1tp}
The tensor product of two type 1 unitary irreducible representations decomposes completely
into type 1 unitary irreducible representations.
\end{thm}

Furthermore, one can easily verify the following.

\begin{thm}
\label{weightsum}
Let $V(\Lambda)$ and $V(\Lambda')$ be irreducible type 1 unitary modules. Then
$V(\Lambda+\Lambda')$ is also irreducible type 1 unitary and occurs in
$V(\Lambda)\otimes V(\Lambda')$.
\end{thm}

It was shown in \cite{ZhaGou1990,GouZha1990} that the type 1 unitary irreducible representations 
are completely characterised by conditions on the highest weight labels:

\begin{thm} \label{gzthm1}
The irreducible highest weight $gl(m|n)$-module $V(\Lambda)$ is type 1 unitary if and only
if $\Lambda$ is real and satisfies
\begin{itemize}
\item[(i)] $(\Lambda+\rho,\varepsilon_m - \delta_n)>0;$ or
\item[(ii)] there exists an odd index $\mu\in\{1,2,\ldots,n\}$ such that
$$
(\Lambda+\rho,\varepsilon_m-\delta_\mu)=0 = (\Lambda,\delta_\mu-\delta_n).
$$
\end{itemize}
\end{thm}

Following Kac \cite{Kac1978}, a finite dimensional irreducible $V(\Lambda)$ is called {\em
typical} if it splits in any finite dimensional module. In other words, if it is a
submodule, then it must occur as a direct summand. If $V(\Lambda)$ is not typical, it is
said to be {\em atypical}. Case (i) of the theorem corresponds to the typical unitary
irreducible modules, whereas case
(ii) corresponds to the atypical ones. Details of the proof of this theorem
can be found in \cite{ZhaGou1990}.

Given a representation $\pi$, its dual representation $\pi^*$ is defined by
\cite{NahmSch1976} 
$$
\pi^*(E_{pq}) = -\left[\pi(E_{pq})\right]^T,
$$
where $T$ denotes supertranspose. Explicitly, in a homogeneous basis 
$\{e_\alpha\}$ of $V$, the supertranspose is defined as
$$
\left( \left[\pi(E_{pq})\right]^T\right)_{\alpha\beta} = (-1)^{[(p)+(q)](\beta)}
\left[\pi(E_{pq})\right]_{\beta\alpha},
$$
where $(\beta)$ denotes the grading of the homogeneous basis vector $e_\beta$, that is,
$(\beta)=0$ (respectively 1) if $e_{\beta}$ is even (respectively odd).

Following \cite{GouZha19902}, we now show that the type 1 and 2 unitary representations
are related by duality in the following sense. If $V$ is a type $\theta$ unitary module
with corresponding representation $\pi$, so that equation (\ref{hermit}) is satisfied,
then the dual representation $\pi^*$ satisfies
\begin{align*}
\left( \left[ \pi^*(E_{pq}) \right]^\dagger \right)_{\alpha\beta} 
&= \left[ \overline{\pi^*(E_{pq})} \right]_{\beta\alpha} \nn\\
&= -\left( \left[ \overline{\pi(E_{pq})} \right]^T \right)_{\beta\alpha }\nn\\
&=  -(-1)^{[(p)+(q)](\alpha)} \left[ \overline{\pi(E_{pq})}  \right]_{\alpha\beta}\nn\\
&=  -(-1)^{[(p)+(q)][(\alpha)+\theta -1]} \left[ \overline{\pi(E_{qp})} \right]_{\beta\alpha}\nn\\
&=  -(-1)^{[(p)+(q)][(\alpha)+(\beta)+\theta -1]} \left( \left[ \pi(E_{qp}) \right]^T \right)_{\alpha\beta}\nn\\
&=  (-1)^{[(p)+(q)][(\alpha)+(\beta)+\theta -1]} \left[ \pi^*(E_{qp})
\right]_{\alpha\beta}.\nn
\end{align*}
Using the fact that $(\alpha)+(\beta) = (p)+(q)$, i.e. the action of an even generator does not change
the grading of a vector in the module, whereas that of an odd generator does, we have 
$$
\left[ \pi^*(E_{pq}) \right]^\dagger  
= (-1)^{[(p)+(q)]\theta} \left[ \pi^*(E_{qp}) \right].
$$
We therefore have the following \cite{GouZha19902}:

\begin{thm}
The dual of a type 1 unitary irreducible representation is a type 2 unitary representation
and vice versa.
\end{thm}

This relationship via duality between the type 1 unitary and type 2 unitary irreducible
representations then allows a complete classification of type 2 unitary irreducible
representations with the following result \cite{ZhaGou1990,GouZha1990}:

\begin{thm}
\label{gzthm2}
The irreducible highest weight $gl(m|n)$-module $V(\Lambda)$ is type 2 unitary if and only
if $\Lambda$ is real and satisfies
\begin{itemize}
\item[(i)] $(\Lambda+\rho,\varepsilon_1 - \delta_1)<0;$ or
\item[(ii)] there exists an even index $k\in\{1,2,\ldots,m\}$ such that
$$
(\Lambda+\rho,\varepsilon_k-\delta_1)=0 = (\Lambda,\varepsilon_k-\varepsilon_1).
$$
\end{itemize}
\end{thm}

%
%

\section{Irreducible covariant tensor and non-tensorial representations} \label{repcat}

The type 1 unitary irreducible representations of $gl(m|n)$ under consideration in the current article
comprise representations which are as follows:
\begin{itemize}
\item[(a)] {\em Covariant tensor}, i.e. those which occur in the tensor
product of a number of copies of the vector representation $V(\varepsilon_1)$ with itself;
\item[(b)] Non-tensorial;
\item[(c)] 
Those which occur in the tensor product of an irreducible representation of class (a) 
with a unitary one dimensional representation. For the purposes of this article, we simply refer to the third class of
representations as {\em extended tensor representations}. 
\end{itemize}

Technically, representations belonging to the
third class may be non-tensorial, but we distinguish them because matrix element formulae for
representations of this general form have not yet been considered in the literature.

The covariant tensor representations are discussed by Gould and Zhang in
the context of their classification scheme \cite{ZhaGou1990} (referred to as
``contravariant'') and also more recently considered by Stoilova and Van der Jeugt 
\cite{StoiVan2010} and Molev \cite{Molev2011} (where they are referred to, though, as ``covariant''). 

The type 1 unitary irreducible representations that we consider in this article
then fall into several distinct cases, and so our matrix element formulae that we develop
here combine several previous works with a genuinely new case (although somewhat
simplistic) into a single formalism as indicated in the following:
\begin{itemize}
\item[(a)] The matrix element formulae of Stoilova and Van der Jeugt \cite{StoiVan2010}
and Molev \cite{Molev2011} corresponding to the irreducible covariant tensor representations, comprising both
typical and atypical representations, but all of which are type 1 unitary;
\item[(b)] The matrix element formulae of Palev \cite{Palev1989}, corresponding to the
so-called {\em essentially typical} representations. 
\item[(c)] The matrix element formulae for irreducible extended tensor representations, 
some of which have not been covered in the existing literature.
\end{itemize}

We point out that there is some overlap between cases (a) and (b) above. In fact, all of
the irreducible typical unitary type 1 representations are either covariant tensor
representations considered by Molev and Stoilova and Van der Jeugt or essentially typical 
and agree with the results of Palev in case (b). 
The typical irreducible representations which are non-tensorial, however, comprise all the
non-tensorial type 1 unitary irreducible representations, except those that occur in case (c).
In the current article, case (c) is genuinely new.
Most importantly, we provide here a universal formalism combining all of these cases. By
contrast, in this context the atypical representations are either covariant tensor or
a tensor product of a covariant tensor representation with a one dimensional
representation.

As a result of the definition of unitary representation, particularly due to the existence
of a positive definite sesquilinear form naturally inherited throughout the canonical
subalgebra embeddings, complete reducibility (in the subalgebra sense) follows immediately
for the type 1 unitary representations (similarly the type 2 unitary representations
also).

A great deal of this discussion can be made explicit by applying the conditions of
atypicality for irreducible type
1 unitary representations given in the classification scheme of Gould and Zhang
\cite{ZhaGou1990,GouZha1990} to the highest weight of the irreducible representation.

Keeping in mind the results of Theorems \ref{type1tp} and \ref{weightsum}, we find that we are able to characterise the irreducible type 1
unitary representations in the following way. We make use of the system of simple roots of $gl(m|n)$ given in equation
(\ref{simpleroots}),
extended by an additional weight ($\varphi_{\overline{n}}$) as follows:
\begin{align*}
\varphi_i &= \varepsilon_i - \varepsilon_{i+1},\ \ 1\leq i <m,\\
\varphi_m &= \varepsilon_m - \delta_1,\\
\varphi_\mu &= \delta_\mu - \delta_{\mu+1},\ \ 1\leq \mu <n,\\
\varphi_{\overline{n}} &= \delta_n.
\end{align*}
We refer to this extended simple root system as $\Phi'$. Note that throughout we use the notation
$\overline{n}$ to indicate that the index is odd (i.e. $\mu=n$), to avoid any ambiguity.

Relative to the inner product $(\Lambda,\Lambda')$
on the weights, we may define a weight basis dual to $\Phi'$ (in the graded sense) which is the analogue of the fundamental dominant weights
for Lie algebras, as 
\begin{align*}
\omega_i &=
(\underbrace{1,1,\ldots,1}_{i},\underbrace{0,0,\ldots,0}_{m-i}|\underbrace{0,0,\ldots,0}_n),\
\ 1\leq i\leq m,
\\
\omega_\mu &= (\underbrace{-1,-1,\ldots,-1}_m|\underbrace{1,1,\ldots,1}_\mu,\underbrace{0,0,\ldots,0}_{n-\mu}),
\ 1\leq \mu\leq n,
\end{align*}
so that
$$
(\omega_i,\varphi_j) = \delta_{ij},\ \ (\omega_\mu,\varphi_\nu) = -\delta_{\mu \nu}, \ \
(\omega_i,\varphi_\nu) = 0 = (\omega_\nu,\varphi_i).
$$
It is straightforward to verify that under the classification scheme outlined in the
previous section, particularly Theorem \ref{gzthm1}, the $\omega_i$ are type 1 unitary, and so is $\omega_{\overline{n}}$.
The $\omega_\mu$ for $1\leq \mu <n$, however, are type 2 unitary. So while we can expand
$\Lambda$ as 
\begin{equation}
\Lambda = \sum_{i=1}^m(\Lambda,\varphi_i)\omega_i -
\sum_{\nu}^n(\lambda,\varphi_\nu)\omega_\nu,
\label{fdwexp}
\end{equation}
in this form it is not clear that $V(\Lambda)$ would be irreducible type 1 unitary. 
This can be achieved, however, by describing such highest
weights in terms of a basis consisting of only type 1 unitary weights. Given that the $\omega_\mu$ are type 2 unitary
for $1\leq \mu<n$, we choose a slightly modified weight basis, denoted
$\{\Omega_r,\varepsilon,\delta\}$, which is no longer dual to
the extended simple root system, but consists of only type 1 unitary highest weights. We
call these the {\em type 1 unitary graded fundamental weights}, defined as
\begin{align*}
\Omega_i &= \omega_i,\ \ 1\leq i< m,\\
\Omega_\mu &= (\mu+1)\omega_m + \omega_\mu,\ \ 1\leq \mu<n,\\
\varepsilon &= \omega_m,\\
\delta &= \omega_{\overline{n}}.
\end{align*}

From equation (\ref{fdwexp}), we can then write 
\begin{equation}
\Lambda = \sum_{i=1}^{m-1}(\Lambda,\varphi_i)\Omega_i 
-\sum_{\nu=1}^{n-1}(\Lambda,\varphi_\nu)\Omega_\nu 
+ \left( (\Lambda,\varphi_m)+\sum_{\nu=1}^{n-1}(\nu+1)(\Lambda,\varphi_\nu) \right)\varepsilon 
-(\Lambda,\varphi_{\overline{n}})\delta.
\label{ugfwexp}
\end{equation}
With $\Lambda = (\Lambda^0_1,\Lambda^0_2,\ldots,\Lambda^0_m\ |\
\Lambda^1_1,\Lambda^1_2,\ldots,\Lambda^1_n),$ we can see that 
$$
(\Lambda,\varphi_i) = \Lambda^0_i - \Lambda^0_{i+1},\ \ i=1,2,\ldots,m-1,
$$
and
$$
(\Lambda,\varphi_\nu) = -\Lambda^1_\nu + \Lambda^1_{\nu+1},\ \ \nu=1,2,\ldots,n-1,
$$
so that $0\leq (\Lambda,\varphi_i)\in\mathbb{Z}$ and $0\leq
-(\Lambda,\varphi_\nu)\in\mathbb{Z}$ since $\Lambda$ is dominant. In other words, the first two of the four
terms given in the expansion of $\Lambda$ in equation (\ref{ugfwexp}) are a
$\mathbb{Z}_+$-linear combination of the subset
$$
\{ \Omega_i,\ \Omega_\mu\ |\ 1\leq i<m,1\leq \mu<n\},
$$
all of which correspond to irreducible type 1 unitary representations that are tensorial. 
Therefore this part of the expansion of $\Lambda$ comprises only tensorial weights and
therefore is itself tensorial.

Concerning the coefficient of $\varepsilon$ in the expansion (\ref{ugfwexp}) of $\Lambda$,
consider the following.

\begin{lemma}
\label{type1var}
The irreducible $gl(m|n)$ module $V(\gamma\varepsilon)$ is type 1 unitary
if and only if
$\gamma=0,1,2,\ldots,n-1$ or $n-1<\gamma\in\mathbb{R}.$
\end{lemma}

\proof{
This follows immediately from the classification result of Theorem \ref{gzthm1}. Note that
in the case $\gamma=0,1,2,\ldots,n-1$, $V(\gamma\varepsilon)$ is atypical and 
$n-1<\gamma\in\mathbb{R}$ corresponds to the case $V(\gamma\varepsilon)$ is typical.
}

The significance of this result is that by Theorem \ref{weightsum} and the expansion
(\ref{ugfwexp}) of $\Lambda$, if $V(\Lambda)$ is irreducible type 1 unitary, then so is
$V(\Lambda+\gamma\varepsilon)$, specifically with $\gamma$ being of the form of the
coefficient of $\varepsilon$ in equation (\ref{ugfwexp}) and satisfying the conditions of
Lemma \ref{type1var}.

In fact, when $\gamma$ (i.e. the coefficient of $\varepsilon$ in (\ref{ugfwexp}))
satisfies the conditions of Lemma \ref{type1var} and only takes integer values, this will
also give rise to a tensorial representation. In the case $\gamma>n-1$ and is non-integer
(and typical), this will correspond to a non-tensorial representation. It is these
representations which include the results of Palev
\cite{Palev1989} that were not treated in the covariant tensor case by Stoilova and Van der Jeugt
\cite{StoiVan2010} or Molev \cite{Molev2011}.

Finally, it is easily seen that 
$$
(\Lambda,\varphi_{\overline{n}}) = -\Lambda^1_n\in \mathbb{R},
$$
so the coefficient of $\delta$ in (\ref{ugfwexp}) may be real in general. Note the
following.

\begin{lemma}
\label{type1del}
For any $\omega\in\mathbb{R}$, the irreducible $gl(m|n)$ module $V(\omega\delta)$ is type
1 unitary and one dimensional.   
\end{lemma}

\proof{
The fact that the module is type 1 unitary (and atypical in fact) is yet another direct consequence of Theorem
\ref{gzthm1}. Using the commutation relations, it is also straightforward to see that the
module is one dimensional for any value $\omega\in\mathbb{R}$.
}

\noindent {\bf Note:} These one dimensional modules are also type 2 unitary as can be seen
from Theorem \ref{gzthm2}. In fact these are the unique irreducible modules that are both
type 1 and type 2 unitary.

The fourth and final term in the expansion (\ref{ugfwexp}) of $\Lambda$ is clearly
relevant to Lemma \ref{type1del}. Theorem \ref{weightsum} then indicates that a highest weight $\Lambda$ with a
non-trivial component of $\delta$ is non-tensorial and occurs in the tensor product of
a one dimensional module (i.e. $V(\omega\delta)$) with either a tensorial representation
(like those considered in \cite{StoiVan2010} and \cite{Palev1989}) 
or a non-tensorial representation, which also includes those considered in \cite{Palev1989}. 

We present a summary of the discussion and results of this section in the following theorem.

\begin{thm}
\label{secsum}
The highest weight $\Lambda$ of an irreducible type 1 unitary $gl(m|n)$ representation is
expressible as 
$$
\Lambda  = \Lambda_0 + \gamma\varepsilon + \omega\delta,
$$
where $\Lambda_0$ is the highest weight of an irreducible tensorial (type 1 unitary)
representation, $\gamma\in \mathbb{R}$ satisfies the conditions of Lemma \ref{type1var},
and $\omega\in\mathbb{R}$.
\end{thm}

The importance of the discussion in this section is that it puts our results of Section
\ref{meg} into
context. In Section \ref{meg}, we determine matrix element formulae for the
$gl(m|n)$ generators on any finite dimensional type 1 unitary representation. Theorem
\ref{secsum} gives us a straightforward characterisation of all irreducible type 1 unitary
irreducible representations relative to the cases (a), (b) and (c) mentioned at the
beginning of this section.
\begin{itemize}
\item[(a)] In the case $\omega=0$ and $\gamma$ an integer, our results will recover
those of Stoilova and Van der Jeugt \cite{StoiVan2010} and a subset of the results of  
Palev \cite{Palev1989} which are also tensorial;
\item[(b)] For $\omega=0$ and non-integer values of
$\gamma$, the representations are non-tensorial, and our results once again include the
work of Palev \cite{Palev1989}. For the case under consideration, i.e. unitary representations, 
these non-tensorial representations are all essentially typical.
\item[(c)] For $\omega\neq 0$, however, our formulae takes into account cases that have
not appeared in the literature. In this case, the representations are generally
non-tensorial, although in a somewhat simplistic sense. They correspond to the extended
tensor representations discussed earlier. For this case, there is some translational invariance
in the sense that the matrix element formulae that we derive in the next section do not
depend on $\omega$. In addition, for distinct values of $\omega$, these representations
will be isomorphic as representations of $sl(m|n)$.
\end{itemize}


\section{Branching rules} \label{branch}

Throughout this paper, where appropriate we use the Gelfand-Tsetlin (GT) basis notation 
where $\lambda_{r,p}$ is the weight label located at the $r$th position in the $p$th row.
The GT patterns for $gl(m|n+1)$ that we consider can be written as 
\begin{equation}
\left|
\begin{array}{cccccccccc}
\lambda_{1,m+n+1} & \lambda_{2,m+n+1} & \cdots & \lambda_{m,m+n+1} & | & 
\lambda_{\bar{1},m+n+1} & \lambda_{\bar{2},m+n+1} & \cdots &
\lambda_{\bar{n},m+n+1} & \lambda_{\overline{n+1},m+n+1} 
\\
\lambda_{1,m+n} & \lambda_{2,m+n} & \cdots & \lambda_{m,m+n} & | &
\lambda_{\bar{1},m+n} & \lambda_{\bar{2},m+n} &\cdots & \lambda_{\bar{n},m+n} &\\
\vdots &  & & & \vdots & &   & \iddots && \\
\lambda_{1,m+1} & \lambda_{2,m+1} & \cdots & \lambda_{m,m+1} & | & \lambda_{\bar{1},m+1} &&&&\\
 &  & \cdots &  &  &  &  &  &  &   \\
\lambda_{1,m} & \lambda_{2,m}& \cdots & \lambda_{m,m} &  &&&&&\\
\vdots &  & \iddots & &  & &  & && \\
\lambda_{1,2} & \lambda_{2,2} & &  & &  & & & &\\
\lambda_{1,1} & & & & & & & & &  
\end{array}
\right)
\label{genGT}
\end{equation}
where each row is a highest weight corresponding to an irreducible representation
permitted by the branching rule for the subalgebra chain
\begin{equation}
gl(m|n+1)\supset gl(m|n)\supset \cdots \supset gl(m|1)\supset gl(m)\supset
gl(m-1)\supset \cdots \supset gl(1).
\label{flag}
\end{equation}

With regards to the branching rules, using the notation above, we first recall the {\em branching conditions}
presented in \cite{GIW1}, which serve as a necessary condition on the $gl(m|p)$ highest
weights occuring in the branching rule of an irreducible $gl(m|p+1)$ highest weight
representation.

\begin{thm} \label{pg1} \cite{GIW1}
For $r\geq m+1$, the following conditions on the dominant weight labels must hold in the pattern
(\ref{genGT}):
\begin{align*}
\lambda_{\mu,r+1}\geq \lambda_{\mu,r}\geq \lambda_{\mu+1,r+1}, & \ \ 1\leq \mu\leq n,\\
\lambda_{i,r+1} \geq \lambda_{i,r}\geq \lambda_{i,r+1}-1, &\ \ 1\leq i\leq m.
\end{align*}
\end{thm}

We also make use of the results of \cite{GouBraHug1989,GouJarBra1990} specific to
$gl(m|1)\supset gl(m)$, adapted to the type 1 unitary representations under
consideration.

\begin{thm} \label{pg2} \cite{GouBraHug1989,GouJarBra1990}
For a unitary type 1 irreducible representation $V(\Lambda)$ of $gl(m|1)$, 
using the notation of (\ref{genGT}),
we have the following conditions on the dominant weight labels: 
\begin{align*}
\lambda_{i,m+1}\geq \lambda_{i,m}\geq \lambda_{i,m+1}-1, &\ \ 1\leq i\leq m-1,\\
\lambda_{m,m+1}\geq \lambda_{m,m}\geq \lambda_{m,m+1}-1, &\ \
\mbox{if }(\Lambda+\rho,\varepsilon_m-\delta_1)>0 \mbox{ (i.e. only if $\Lambda$ typical),}\\
\lambda_{m,m}= \lambda_{m,m+1}, &\ \
\mbox{if }(\Lambda+\rho,\varepsilon_m-\delta_1)=0 \mbox{ (i.e. only if $\Lambda$
atypical).}\\
\end{align*}
\end{thm}

For the general $gl(m|n+1)$ branching rule, 
we have the following result.

\begin{thm} \label{mainbranchingrule}
For a unitary type 1 irreducible $gl(m|n+1)$ representation, the basis vectors can be
expressed in the form (\ref{genGT}), with the following conditions on the dominant weight labels:
\begin{itemize}
\item[(1)] 
For $r\geq m+1$, \\
$\lambda_{\mu,r+1}\geq \lambda_{\mu,r}\geq \lambda_{\mu+1,r+1},$ $1\leq \mu\leq n, $\\ 
$\lambda_{i,r+1} \geq \lambda_{i,r}\geq \lambda_{i,r+1}-1,$ $1\leq i\leq m$, \\
(i.e result of Theorem \ref{pg1});
\item[(2)] 
$\lambda_{i,m+1}\geq \lambda_{i,m}\geq \lambda_{i,m+1}-1,$ $1\leq i\leq m-1,$\\
$\lambda_{m,m+1}\geq \lambda_{m,m}\geq \lambda_{m,m+1}-1,$
if $(\Lambda+\rho,\varepsilon_m-\delta_1)>0$ ($\Leftrightarrow$ only if $\Lambda$
typical),\\
$\lambda_{m,m}= \lambda_{m,m+1},$ 
if $(\Lambda+\rho,\varepsilon_m-\delta_1)=0$ ($\Leftrightarrow$ only if $\Lambda$
atypical),\\
(i.e. result of Theorem \ref{pg2});
\item[(3)] For $1\leq j\leq m$, \\
$\lambda_{i+1,j+1}\geq \lambda_{i,j}\geq \lambda_{i,j+1}$ \\
(i.e. the usual $gl(j)$ branching rules);
\item[(4)] For each $r$ such that $m+1\leq r\leq m+n+1$, the $r$th row in (\ref{genGT}) must
correspond to a type 1 unitary highest $gl(m|r)$ weight, and for each $j$ such that
$1\leq j\leq m$, the $j$th row in (\ref{genGT}) must correspond to a highest $gl(j)$
weight. 
\end{itemize}
\end{thm}


In the case of covariant tensor representations, it is easily verified that our branching rules 
coincide with those already presented by Stoilova and Van der Jeugt in \cite{StoiVan2010}. 
Otherwise, the representation must be non-tensorial and essentially typical (modulo a one
dimensional representation with highest weight of the form
$\omega\delta$), and our branching rules are the 
full branching conditions given in Theorem \ref{pg1},
which furthermore coincide with the branching rules given by Palev for the essentially
typical representations \cite{Palev1989,PaStVa1994}. For a proof of the branching rule for
essentially typical representations, see Appendix A.

%
%

\section{Matrix elements of generators}\label{meg}

We now recall some of the definitions and results from our article \cite{GIW1} which will
be used to derive the matrix element formulae of the current article. Let $Q[r]$ and
$\bar{Q}[r]$ ($1\leq r \leq m+n+1$) denote the projection operators for $gl(m|n+1)$ which
are analogues of the $gl(m|n)$ projections $P[s]$ and $\bar{P}[s]$ ($1\leq s\leq m+n$) respectively. The
$gl(m|n+1)$ characteristic roots are denoted $\beta_r$ and $\bar{\beta}_r$ which are
the respective counterparts to the $gl(m|n)$ characteristic roots $\alpha_s$ and
$\bar{\alpha}_s.$ We also let $\psi^p$ and $\phi_p$ denote the odd $gl(m|n)$
vector and contragredient vector operators respectively defined by
\begin{align}
\psi^p &= (-1)^{(p)}E_{p,m+n+1}={\cal B}^p_{\ m+n+1},\nn\\
\phi_p &= (-1)^{(p)}E_{m+n+1,p} = -(-1)^{(p)}{\cal B}^{m+n+1}_{\ \ \ \ \ \ \ p}.\nn
\end{align}

Let $\alpha_r$ denote characteristic roots corresponding to the $gl(m|n)$-module $V(\Lambda)$ and
$\beta_r$ denote the characteristic roots of the $gl(m|n+1)$-module $V(\tilde{\Lambda})$ such
that $V(\Lambda)\subseteq V(\tilde{\Lambda})$ as a $gl(m|n)$-module. 
In \cite{GIW1}, we found that the betweenness conditions imply, for $1\leq i\leq m$,
\begin{align}
\beta_i = \left\{ \begin{array}{rl} \alpha_i,& \tilde{\Lambda}_i = 1+\Lambda_i\\
                                    \alpha_i - 1,& \tilde{\Lambda}_i = \Lambda_i 
\end{array} \right.
\nn
\end{align}
which leads us to make use of the following index sets: 
\begin{align}
I_0 &=  \{ 1\leq i\leq m\ |\ \alpha_i=\beta_i\},\nn\\
\bar{I}_0 &=  \{ 1\leq i\leq m\ |\ \alpha_i=1+\beta_i\},\nn\\
I_1 &= \{ 1\leq\mu\leq n\},\nn\\
I &= I_0\cup I_1,\nn\\
I'&= \bar{I}_0\cup I_1,\nn\\
\tilde{I} &= I\cup \{m+n+1\},\nn\\
\tilde{I}' &= I'\cup \{m+n+1\}.\nn
\end{align}

The (even) $gl(m|n)$ invariants
%
%
\begin{align}
c_r = Q[r]^{m+n+1}_{\ \ \ \ \ \ \ m+n+1},\ \ \bar{c}_r = \bar{Q}[r]_{m+n+1}^{\ \ \ \ \ \ \ m+n+1},\ \ 1\leq r\leq m+n+1
\label{equ5.2}
\end{align}
can be shown to satisfy
%
%
\begin{align}
Q[r]^p_{\ m+n+1} &= \sum_{q\in I} \psi[q]^p(\beta_r-\alpha_q-(-1)^{(q)})^{-1}c_r,\
\ r\in \tilde{I}, 
\nn\\
\bar{Q}[r]_p^{\ m+n+1} &= -\sum_{q\in I'}\phi[q]_p(\bar{\beta}_r- \bar{\alpha}_q -
(-1)^{(q)})^{-1} \bar{c}_r,\ r\in \tilde{I}'
\nn
\end{align}
and have eigenvalues given by
%
%
\begin{align}
c_r &= \prod_{k\in\tilde{I},k\neq r} (\beta_r - \beta_k)^{-1}\prod_{s\in I}(\beta_r -
\alpha_s-(-1)^{(s)}),\ \ r\in \tilde{I},
\label{equ5.10}\\
c_r &= 0, \ \ r\notin \tilde{I},\nn\\
\bar{c}_r &= \prod_{k\in \tilde{I}',k\neq r} \left(\bar{\beta}_r - \bar{\beta}_k\right)^{-1}\prod_{s\in
I'} \left(\bar{\beta}_r - \bar{\alpha}_s - (-1)^{(s)}\right),\ \ r\in \tilde{I}',
\label{equ5.12}\\
\bar{c}_r &= 0,\ \ r\notin \tilde{I}'.\nn
\end{align}
It was also argued in \cite{GIW1} that we may define invariants $\delta_r$ and $\bar{\delta}_r$ satisfying
%
%
\begin{align}
(-1)^{(q)}\psi[r]^p\phi[r]_q &= \delta_r P[r]^p_{\ q},\ \ r\in
I',\label{equ5.19a}\\
\phi[r]_p \psi[r]^q &= \bar{\delta}_r \bar{P}[r]_p^{\ q},\ \ r\in I,
\label{equ5.19b}
\end{align}
with eigenvalues given by
\begin{align}
\delta_r &= \prod_{q\in I, q\neq r} (\alpha_q - \alpha_r +
(-1)^{(q)})^{-1}\prod_{s\in \tilde{I}} (\beta_s-\alpha_r), \ \ r\in I',
\label{deltaform}\\
\delta_r &= 0,\ \ r\notin I',\nn\\
\bar{\delta}_r &=  -\prod_{q\in I', q\neq r} (\bar{\alpha}_q - \bar{\alpha}_r +
(-1)^{(q)})^{-1}\prod_{s\in \tilde{I}'} (\bar{\beta}_s-\bar{\alpha}_r), \ \ r\in I,
\label{deltabarform} \\
\bar{\delta}_r &= 0, r\notin{I}.\nn
\end{align}
Taking $p=q=m+n$ in equations (\ref{equ5.19a}) and (\ref{equ5.19b}) we obtain the equations
\begin{align}
\psi[r]^{m+n}\phi[r]_{m+n} &= -\delta_r P[r]^{m+n}_{\ m+n} \label{equ6.1a}\\
\phi[r]_{m+n}\psi[r]^{m+n} &= \bar{\delta}_r \bar{P}[r]_{m+n}^{\ m+n} \label{equ6.1b}
\end{align}
where we note that
$$
P[r]^{m+n}_{\ m+n},\ \ \bar{P}[r]_{m+n}^{\ m+n}
$$
are the $gl(m|n)$ analogues of the invariants $c_r$ and $\bar{c}_r$ of equation
(\ref{equ5.2}) which may similarly be expressed in terms of the $gl(m|n)$ and $gl(m|n-1)$
characteristic roots in accordance with equations (\ref{equ5.10}) and (\ref{equ5.12})
respectively. Thus the operators on the right hand side of equations (\ref{equ6.1a}) and
(\ref{equ6.1b}) may be simply evaluated as a rational polynomial function in the
representation labels of $gl(m|n+1)$, $gl(m|n)$ and $gl(m|n-1)$.

In the case of unitary representations with
$$
\left( \psi[r]^p \right)^\dagger = \phi[r]_p,
$$
we note that equations (\ref{equ6.1a}) and (\ref{equ6.1b}) determine the square of the matrix elements of
$\phi_{m+n}$ and $\psi_{m+n}$ respectively. Thus we take the formulae arising from 
equations (\ref{equ6.1a}) and (\ref{equ6.1b}) to determine the matrix elements. 

Before this can be done, we need a method for determining the non-elementary generator
matrix elements of $\phi_p$ and $\psi^p$ for $p<m+n$. 
To this end, we adopt an approach that applies the characteristic identity
(\ref{veccharid}) that was first used in \cite{Gould1992}.

We first note that it is straightforward to verify that 
\begin{align*}
(\beta_r-\alpha_p)^{-1}\psi[p] &= \psi[p](\beta_r-\alpha_p - (-1)^{(p)})^{-1},\\
\phi[p](\beta_r-\alpha_p)^{-1} &= (\beta_r-\alpha_p-(-1)^{(p)})^{-1}\phi[p].
\end{align*}
Let ${\cal B}$ denote the characteristic matrix for $gl(m|n+1)$, being the analogue of
${\cal A}$, defined earlier in equation (\ref{charmat}). Using the projection operators 
$$
Q[r] = \prod_{k\neq r}^{m+n+1}\left( \frac{{\cal B}-\beta_k}{\beta_r-\beta_k} \right),
$$
the characteristic identity can be expressed as
$$
{\cal B}^s_{\ u}Q[r]^u_{\ t} = \beta_rQ[r]^s_{\ t},\ \ 1\leq r,s,t,u\leq m+n+1,
$$
which then leads to
$$
{\cal B}^s_{\ m+n+1}Q[r]^{m+n+1}_{\ \ \ \ \ \ \ t}+{\cal B}^s_{\ {u}}Q[r]^{{u}}_{\ t} =
\beta_rQ[r]^s_{\ t},
$$
with $u$ being summed from 1 to $m+n$.
Note that we are using the summation convention where we sum over repeated indices from 1
up to either $m+n$ or $m+n+1$ depending on the context. If it is not obvious, we make an
explicit note.

If we further restrict the $s$ index in the previous equation up to $m+n$, we may then write
$$
\psi^{ {s}}Q[r]^{m+n+1}_{\ \ \ \ \ \ \ t} = (\beta_r-{\cal A})^{ {s}}_{\
 {u}}Q[r]^{ {u}}_{\ t}.
$$
Multiplying from the left by $P[u]$ gives
\begin{align}
P[u]^{ {s}}_{\  {w}}\psi^{ {w}}Q[r]^{m+n+1}_{\ \ \ \ \ \ \ t}
&=
P[u]^{ {s}}_{\  {x}}(\beta_r-{\cal A})^{ {x}}_{\  {v}}Q[r]^{ {v}}_{\
t}\nn\\
\Rightarrow\ \ 
\psi[u]^{ {s}}Q[r]^{m+n+1}_{\ \ \ \ \ \ \ t}
&= 
(\beta_r - \alpha_u)P[u]^{ {s}}_{\  {v}}Q[r]^{ {v}}_{\ t}\nn\\
\Rightarrow 
(\beta_r - \alpha_u)^{-1}  \psi[u]^{ {s}}Q[r]^{m+n+1}_{\ \ \ \ \ \ \ t}
&= 
P[u]^{ {s}}_{\  {v}}Q[r]^{ {v}}_{\ t}\nn\\
\Rightarrow \ \ \psi[u]^{ {s}}(\beta_r - \alpha_u-(-1)^{(u)})^{-1}  Q[r]^{m+n+1}_{\ \ \ \ \ \ \ t}
&= 
P[u]^{ {s}}_{\  {v}}Q[r]^{ {v}}_{\ t}.
\label{refequ1}
\end{align}
Alternatively, the characteristic identity may be written
$$
Q[r]^s_{\ u}{\cal B}^u_{\ t} = Q[r]^s_{\ t}\beta_r.
$$
Setting $s=m+n+1$ and restricting $t$ to values up to $m+n$ leads to
\begin{align*}
Q[r]^{m+n+1}_{\ \ \ \ \ \ \ v}{\cal B}^v_{\ t} + Q[r]^{m+n+1}_{\ \ \ \ \ \ \ m+n+1}{\cal B}^{m+n+1}_{\ \ \ \ \ \ \ t} &= Q[r]^{m+n+1}_{\ \ \ \ \ \ \ t}\beta_r\\
\Rightarrow\ \ 
Q[r]^{m+n+1}_{\ \ \ \ \ \ \  {v}}{\cal B}^{ {v}}_{\  {t}} -(-1)^{(t)} c_r\phi_t &= Q[r]^{m+n+1}_{\ \ \ \ \ \ \  {t}}\beta_r\\
\Rightarrow\ \ 
-(-1)^{(t)}c_r\phi_{ {t}} = Q[r]^{m+n+1}_{\ \ \ \ \ \ \  {v}}(\beta_r-{\cal A})^{ {v}}_{\ {t}}.
\end{align*}
Multiplying on the right by $P[u]^t_{\ w}(-1)^{(w)}$ then summing over $t$ gives
\begin{align*}
-(-1)^{(t)+(w)}c_r\phi_{ {t}}P[u]^{ {t}}_{\  {w}} &= (-1)^{(w)}Q[r]^{m+n+1}_{\ \ \ \ \ \ \ {v}}(\beta_r-{\cal A})^{ {v}}_{\  {t}}P[u]^{ {t}}_{\  {w}}\\
\Rightarrow\ \ -c_r\phi[u]_{ {w}} &= (-1)^{(w)}Q[r]^{m+n+1}_{\ \ \ \ \ \ \  {v}}P[u]^{ {v}}_{\  {w}}(\beta_r-\alpha_u)\\
\Rightarrow\ \ -c_r\phi[u]_{ {w}}(\beta_r-\alpha_u)^{-1} &= (-1)^{(w)}Q[r]^{m+n+1}_{\ \ \ \ \ \ \  {v}}P[u]^{ {v}}_{\  {w}}\\
\Rightarrow\ \ -(-1)^{(w)}c_r(\beta_r-\alpha_u-(-1)^{(u)})^{-1}\phi[u]_{ {w}} &= Q[r]^{m+n+1}_{\ \ \ \ \ \ \  {v}}P[u]^{ {v}}_{\  {w}}.
\end{align*}
Multiplying equation (\ref{refequ1}) on the right by $P[u]$ gives
\begin{align*}
\psi[u]^s(\beta_r-\alpha_u-(-1)^{(u)})^{-1}Q[r]^{m+n+1}_{\ \ \ \ \ \ \ v}P[u]^v_{\ w}
&=
P[u]^s_{\ t}Q[r]^t_{\ v}P[u]^v_{\ w}\\
\Rightarrow\ \ -(-1)^{(w)}\psi[u]^{ {s}}
(\beta_r-\alpha_u-(-1)^{(u)})^{-1}(\beta_r-\alpha_u-(-1)^{(u)})^{-1} c_r \phi[u]_{ {w}}
&= P[u]^{ {s}}_{\  {t}}Q[r]^{ {t}}_{\  {v}}P[u]^{ {v}}_{\  {w}} .
\end{align*}
We now must apply the shift operators $\psi[u]^s$ and $\phi[u]_{ {w}}$ in the above expression. Three cases must be considered:

For odd $u$ we move $\psi[u]^s$ to the right to give 
\begin{align*}
-(-1)^{(w)}(\beta_r-\alpha_u)^{-1} \psi[u]^{
{s}} (\beta_r-\alpha_u-(-1)^{(u)})^{-1} c_r\phi[u]_{ {w}}
&= P[u]^{ {s}}_{\  {t}}Q[r]^{ {t}}_{\  {v}}P[u]^{ {v}}_{\  {w}}\\
\Rightarrow \ \ -(-1)^{(w)}(\beta_r-\alpha_u)^{-1}(\beta_r-\alpha_u-(-1)^{(u)})^{-1} c_r \psi[u]^{
{s}}\phi[u]_{ {w}}
&= P[u]^{ {s}}_{\  {t}}Q[r]^{ {t}}_{\  {v}}P[u]^{ {v}}_{\  {w}},
\end{align*}
where in the last step we have used the fact that
$(\beta_r-\alpha_u-(-1)^{(u)})^{-1}c_r$ is independent of $u$ (for odd $u$) and so commutes with
$\psi[u]^{ {s}}$.
Equation (\ref{equ5.19a}) then leads to
\begin{align}
\rho_{ru} P[u]^p_{\ q}
&= 
(P[u]Q[r]P[u])^p_{\ q} 
\nn
\end{align}
where
\begin{align}
\rho_{ru} = -
(\beta_r-\alpha_u-(-1)^{(u)})^{-1}(\beta_r-\alpha_u)^{-1}c_r\delta_u, \ \ \ (u) = 1.
\nn
\end{align}
Note that the above expression may be used in the $gl(n)$ case by setting $(u) = 0$ when shifting labels at the subalgebra level $m$ or lower.

For even $u$ and $u \neq r$ the invariant $c_r$ will have two extra terms when located between $\psi[u]^s$ and $\phi[u]_w$ due to index set considerations. We then have
\begin{align*}
&-(-1)^{(w)}\psi[u]^{ {s}}
(\beta_r-\alpha_u-(-1)^{(u)})^{-1}(\beta_r-\alpha_u-(-1)^{(u)})^{-1}c_r\phi[u]_{ {w}} \\
&=  -(-1)^{(w)}\psi[u]^{ {s}}
(\beta_r-\alpha_u-(-1)^{(u)})^{-1}(\beta_r-\alpha_u-(-1)^{(u)})^{-1} (\beta_r - \beta_u)^{-1} (\beta_r - \alpha_u - (-1)^{(u)}) \phi[u]_{ {w}} c_r \\
&=  -(-1)^{(w)}\psi[u]^{ {s}}
(\beta_r-\alpha_u-(-1)^{(u)})^{-1} (\beta_r - \beta_u)^{-1} \phi[u]_{ {w}} c_r \\
&=  -(-1)^{(w)}
(\beta_r-\alpha_u)^{-1} (\beta_r - \beta_u)^{-1} \psi[u]^{ {s}}\phi[u]_{ {w}} c_r \\
&=  -(-1)^{(w)}
(\beta_r-\alpha_u)^{-1} (\beta_r - \alpha_u + 1)^{-1} \psi[u]^{ {s}}\phi[u]_{ {w}} c_r
\end{align*}
since $u \in I'$, which gives
$$
\rho_{ru} = - (\beta_r-\alpha_u)^{-1} (\beta_r - \alpha_u + 1)^{-1} c_r \delta_u ~~~~~ ,(u) = 0, r \neq u.
$$
Finally, for even $u$ and $u = r$ we have
\begin{align*}
&-(-1)^{(w)}\psi[u]^{ {s}}
(\beta_r-\alpha_u-(-1)^{(u)})^{-1}(\beta_r-\alpha_u-(-1)^{(u)})^{-1}c_r\phi[u]_{ {w}} \\
&=  -(-1)^{(w)}\psi[u]^{ {s}}
(\beta_r-\alpha_u-(-1)^{(u)})^{-1}(\beta_r-\alpha_u-(-1)^{(u)})^{-1} (\beta_r - \alpha_u - (-1)^{(u)}) \phi[u]_{ {w}} c_r \\
&=  -(-1)^{(w)}\psi[u]^{ {s}}
(\beta_r-\alpha_u-(-1)^{(u)})^{-1} \phi[u]_{ {w}} c_r \\
&=  -(-1)^{(w)}
(\beta_r-\alpha_u)^{-1} \psi[u]^{ {s}} \phi[u]_{ {w}} c_r \\
&=  (-1)^{(w)} \psi[u]^{ {s}} \phi[u]_{ {w}} c_r ~~\hbox{}
\end{align*}
since $r=u$ and $u \in I'$, which gives
$$
\rho_{uu} = c_u \delta_u ~~~~~ ,(u) = 0.
$$
Combining the above three cases we have
\begin{align*}
\rho_{ru} P[u]^p_{\ q}
&= 
(P[u]Q[r]P[u])^p_{\ q} 
\end{align*}
where
\begin{align*}
\rho_{ru} &= -(\beta_r-\alpha_u + 1)^{-1}(\beta_r-\alpha_u)^{-1}c_r\delta_u, \ \ \ (u) = 1, \\
\rho_{ru} &= -(\beta_r-\alpha_u + 1)^{-1}(\beta_r-\alpha_u)^{-1}c_r\delta_u, \ \ \ (u) = 0,
u \neq r, \\
\rho_{uu} &= c_u\delta_u, \ \ \ (u) = 0, \\
\rho_{ru} &= (\beta_r-\alpha_u - 1)^{-1}(\beta_r-\alpha_u)^{-1}c_r\delta_u, \ \ \ gl(m)
\hbox{~case,} 
\end{align*}
is a $gl(m|n)$ invariant whose eigenvalues in fact determine the square of $gl(m|n+1):gl(m|n)$
reduced vector Wigner coefficients.
We note that the invariants $\rho_{ru}$ are only non-vanishing
when $r\in \tilde{I}$ and $u\in I'$. Similarly for the adjoint projectors we have
\begin{align}
(\bar{P}[u]\bar{Q}[r]\bar{P}[u])_p^{\ q} &= \bar{\rho}_{ru}\bar{P}[u]_p^{\ q}
\label{equ6.3}
\end{align}
where the $\bar{\rho}_{ru}$ is a $gl(m|n)$ invariant operator given by
\begin{align}
\bar{\rho}_{ru} &= (\bar{\beta}_r-\bar{\alpha}_u +
1)^{-1}(\bar{\beta}_r-\bar{\alpha}_u)^{-1}\bar{c}_r\bar{\delta}_u, \ \ \ (u) = 1, \nn\\
\bar{\rho}_{ru} &= (\bar{\beta}_r-\bar{\alpha}_u +
1)^{-1}(\bar{\beta}_r-\bar{\alpha}_u)^{-1}\bar{c}_r\bar{\delta}_u \ \ \ (u) = 0, u \neq r,
\nn\\
\bar{\rho}_{uu} &= \bar{c}_u \bar{\delta}_u, \ \ \ (u) = 0, \nn\\
\bar{\rho}_{ru} &= (\bar{\beta}_r-\bar{\alpha}_u -
1)^{-1}(\bar{\beta}_r-\bar{\alpha}_u)^{-1}\bar{c}_r \bar{\delta}_u \ \ \ gl(m)
\hbox{~case,} 
\label{equ6.4}
\end{align}
whose eigenvalues determine the square of certain $gl(m|n+1):gl(m|n)$ reduced Wigner
coefficients. As in the Lie algebra case \cite{Gould1981} the above equations are all we need to
determine the matrix elements of the $gl(m|n+1)$ generators. We note that
$\bar{\rho}_{ru}$ in equation (\ref{equ6.4}) is non-vanishing only when $r\in \tilde{I}'$
and $u\in I$.

{\bf Remark:} \label{remarkref} 
We make the observation that the expressions for $\bar{c}_r$ and $\bar{\delta}_r$, given in
equations (\ref{equ5.12}) and (\ref{deltabarform}) respectively, take exactly the
same form (up to an overall sign in the case of $\bar{\delta}_r$) as their counterparts
$c_r$ and $\delta_r$ of equations (\ref{equ5.10}) and (\ref{deltaform}) respectively.
Specifically, the characteristic roots $\bar{\alpha}_q$ and $\bar{\beta}_q$ are merely
substituted for $\alpha_q$ and $\beta_q$ respectively, and the index sets $I'$ and 
$\tilde{I}'$ are substituted for $I$ and $\tilde{I}$ over the products. Clearly this
symmetry also extends to the expressions for $\rho_{ru}$ and $\bar{\rho}_{ru}$.

\subsection{Matrix element formulae}

In general the $gl(m|n+1)$ generators
$$
\psi^p = (-1)^{(p)} E_{p,m+n+1},\ \ \phi_p =
(-1)^{(p)}E_{m+n+1,p}
$$
may be resolved into a sum of {\em simultaneous shift} components
\begin{align}
\psi^p &= \sum_u \psi[u_{m+n}u_{m+n-1}\ldots u_{p+1}u_p]^p, \label{equ6.5a}\\
\phi_p &= \sum_u \phi[u_{m+n}u_{m+n-1}\ldots u_{p+1}u_p]_p,
\label{equ6.5b}
\end{align}
where the summations in equations (\ref{equ6.5a}) and (\ref{equ6.5b}) are over all
allowable shift components $u_r$ for the subalgebra $gl(m|r-m)$ (in the case $r>m$) or the
subalgebra $gl(r)$ in the case $r\leq m$. In other words, $u_r$ takes all allowable shift
values in the range $1,2,\ldots,r$. The simultaneous shift components of equations
(\ref{equ6.5a}) and (\ref{equ6.5b}) may be defined recursively according to
\begin{align}
\psi[u_{m+n}u_{m+n-1}\ldots u_{q+1}u_q]^p &= 
\sum_{s=1}^q \psi[u_{m+n}\ldots u_{q+1}]^s\bar{P}[u_q]_s^{\ p}\nn\\
&= \sum_{s=1}^q P[u_q]^p_{\ s}\psi[u_{m+n}\ldots u_{q+1}]^s,\ \ 1\leq
p\leq q,\nn\\
\phi[u_{m+n}u_{m+n-1}\ldots u_{q+1}u_q]_p &= 
\sum_{s=1}^q (-1)^{(p)+(s)}\phi[u_{m+n}\ldots u_{q+1}]_s P[u_q]^s_{\ p}\nn\\
&= \sum_{s=1}^q \bar{P}[u_q]_p^{\ s}\phi[u_{m+n}\ldots u_{q+1}]_s,\ \ 1\leq
p\leq q.\nn
\end{align}
Thus we obtain, by repeated application of equations (\ref{equ6.1a}) and (\ref{equ6.3})
\begin{align}
(-1)^{(p)}\psi[u_{m+n}\ldots u_p]^p \phi[u_{m+n}\ldots u_p]_p
&=
\delta_{u_{m+n}} P[u_p u_{p+1}\ldots u_{m+n-1}u_{m+n}u_{m+n-1}\ldots u_{p}]^p_{\ p}
\nn\\
&= \delta_{u_{m+n}}c_{u_p}\prod_{s=p+1}^{m+n} \rho_{u_s,u_{s-1}}
\label{equ6.6}
\end{align}
and similarly
\begin{align}
\phi[u_{m+n}\ldots u_{p}]_p\psi[u_{m+n}\ldots u_p]^p
=
\bar{\delta}_{u_{m+n}}\bar{c}_{u_p}\prod_{s=p+1}^{m+n}\bar{\rho}_{u_s,u_{s-1}}
\label{equ6.7}
\end{align}
which is the required generalisation of equation (\ref{equ6.1b}). 
We remark that in
equations (\ref{equ6.6}) and (\ref{equ6.7}) above, the invariants
$\rho_{u_s,u_{s-1}}$, $c_{u_{s}}$ and $\delta_{u_{s-1}}$ are expressible in
terms of the characteristic roots of $gl(m|s-m)$ and $gl(m|s-m-1)$ by analogy with
our previous formulae (or the Lie algebras $gl(s)$ and $gl(s-1)$ when $s\leq
m$).
In the case of unitary representations, equations (\ref{equ6.6}) and (\ref{equ6.7}) determine
the matrix elements of the $gl(m|n+1)$ generators $\phi_p$ and $\psi^p$
respectively. 

We now give closed form expressions for the matrix elements of the
generators $E_{l,p+1}$ and $E_{p+1,l}$ $(1 \leq l \leq p)$. Once again using the
Gelfand-Tsetlin (GT) basis notation 
with the label $\lambda_{r,p}$ located at the $r$th position in the $p$th row. 
The matrix of $E_{p+1,p+1}$ is diagonal with the entries
\begin{align}
\sum_{r=1}^{p+1} \lambda_{r,p+1} - \sum_{r=1}^{p} \lambda_{r,p}. \nn
\end{align}
We consider a fixed GT pattern denoted by $| \lambda_{q,s} \rangle$ and proceed by first obtaining 
the matrix elements of the elementary generators $E_{p,p+1}$ and $E_{p+1,p}$. 

We first resolve $E_{p,p+1}$ into its shift components, which gives
\begin{align}
E_{p,p+1} | \lambda_{q,s} \rangle &= \sum_{r=1}^p (-1)^{(p)}\psi[r]^p   | \lambda_{q,s} \rangle \nn\\
&= \sum_{r=1}^p N^p_r ( \lambda_{q,p+1} ; \lambda_{q,p} ; \lambda_{q,p-1}) 
| \lambda_{q,s} + \Delta_{r,p} \rangle, \nn
\end{align}
where $| \lambda_{q,s} + \Delta_{r,p} \rangle$ indicates the GT pattern obtained from 
$| \lambda_{q,s} \rangle$ by increasing the label $\lambda_{r,p}$ by one unit and leaving 
the remaining labels unchanged. 

{\bf Remark:} We adopt the convention throughout the article that $| \lambda_{q,s} +
\Delta_{r,p} \rangle$ is identically zero if the branching rules are not satisfied. In
other words, $| \lambda_{q,s} + \Delta_{r,p} \rangle$ does not form an allowable GT
pattern. In such a case the matrix element is understood to be identically zero. 

Since the shift operators satisfy the Hermiticity condition
\begin{align}
\phi[r]_p = \left[ \psi[r]^p \right]^\dagger \nn
\end{align}
then we may use equation (\ref{equ6.1b}) to express the matrix elements $N^p_r$ as 
\begin{align}
N^p_r ( \lambda_{q,p+1} ; \lambda_{q,p} ; \lambda_{q,p-1}) = 
\langle \lambda_{q,s} | \bar{\delta}_{r,p} \bar{c}_{r,p} |  \lambda_{q,s} \rangle^{1/2} \nn
\end{align}
where $\bar{\delta}_{r,p}$ and $\bar{c}_{r,p}$ are either invariants of the $gl(m|p-m)$ subalgebra 
for $m < p \leq m+n$ or invariants of the $gl(p)$ subalgebra for $0 < p \leq m$.

The matrix element $N^p_r$ has an undetermined sign (or phase factor). However, the Baird and Biedenharn 
convention sets the phases of the matrix elements of the elementary generators $E_{p,p+1}$ to be real
and positive - we will follow \cite{Gould1981} and adopt this convention. Matrix element phases 
for the non-elementary generators will be discussed later in this section.

Expressions for the eigenvalues of the invariants $\bar{c}_r$ and $\bar{\delta}_r$
independent of the index sets were given in \cite{GIW1}, namely  
\begin{align}
\bar{c}_i &= (\beta_i-\alpha_i)\prod_{k\neq i}^m\left( \frac{\beta_k-\beta_i-1}{\alpha_k-\beta_i-1} \right)
\prod_{\nu=1}^{n+1}(\beta_\nu - \beta_i-2)^{-1}
\prod_{\nu=1}^n(\alpha_\nu-\beta_i-2),\ \ 1\leq i\leq m,\nn\\
\bar{c}_\mu &= \prod_{k=1}^m\left( \frac{\beta_k-\beta_\mu+1}{\alpha_k-\beta_\mu+1} \right)
\prod_{\nu\neq\mu}^{n+1}(\beta_\nu-\beta_\mu)^{-1}
\prod_{\nu=1}^n(\alpha_\nu -\beta_\mu),\ \ 1\leq\mu\leq n+1,\nn
\end{align}
and
\begin{align}
\bar{\delta}_i &= (\beta_i-\alpha_i+1) \prod_{k\neq
i}^m\left(\frac{\alpha_k-\alpha_i}{\beta_k-\alpha_i}  \right)
\prod_{\nu=1}^{n+1}(\beta_\nu-\alpha_i-1)
\prod_{\nu=1}^n(\alpha_\nu-\alpha_i-1)^{-1},\ \ 1\leq i\leq m,\nn\\
\bar{\delta}_\mu &= -\prod_{k=1}^m\left(
\frac{\alpha_k-\alpha_\mu+2}{\beta_k-\alpha_\mu+2} \right)
\prod_{\nu=1}^{n+1}(\beta_\nu-\alpha_\mu+1)
\prod_{\nu\neq\mu}^n(\alpha_\nu-\alpha_\mu+1)^{-1},\ \ 1\leq\mu\leq n.\nn
\end{align}

To consider these invariants as invariants of the $gl(m|p-m)$ subalgebra we need to extend our notation. 
For the $\bar{c}$ equation we carry out the replacements $\beta_a \rightarrow \alpha_{a,p}$  	
and $\alpha_a \rightarrow \alpha_{a,p-1}$. Similarly, for the $\bar{\delta}$ equation 
we have $\beta_a \rightarrow \alpha_{a,p+1}$ and $\alpha_a \rightarrow \alpha_{a,p}$.
Similarly, a subscript has been added to the index set notation to indicate the subalgebra
level of the roots being compared so that for $p \geq m$ we have
\begin{align}
I_{0,p} &=  \{ 1\leq i\leq m\ |\ \alpha_{i,p}=\alpha_{i,p+1}\},\nn\\
\bar{I}_{0,p} &=  \{ 1\leq i\leq m\ |\ \alpha_{i,p}=1+\alpha_{i,p+1}\},\nn\\
I_{1,p} &= \{ 1\leq\mu\leq p-m\},\nn\\
I_p &= I_{0,p} \cup I_{1,p},\nn\\
I'_p&= \bar{I}_{0,p} \cup I_{1,p},\nn\\
\tilde{I_p} &= {I_p} \cup \{p+1\},\nn\\
\tilde{I_p}' &= I'_p \cup \{p+1\}.\nn
\end{align}

We then obtain
\begin{align}
N^p_i =  \Bigg[ &\prod_{k\neq i = 1}^m \left( \frac
{(\alpha_{k,p} - \alpha_{i,p} -1)(\alpha_{k,p} - \alpha_{i,p} )}
{(\alpha_{k,p-1} - \alpha_{i,p} - 1)(\alpha_{k,p+1} - \alpha_{i,p} )}
\right) \nn\\
\times 
&\left( \frac
{ \prod^{p-m-1}_{\nu = 1} (\alpha_{\nu,p-1} - \alpha_{i,p} - 2) \prod_{\nu=1}^{p-m+1} (\alpha_{\nu,p+1} - \alpha_{i,p} - 1)}
{ \prod^{p-m}_{\nu = 1} (\alpha_{\nu,p} - \alpha_{i,p} - 2 )({\alpha_{\nu,p} - \alpha_{i,p} - 1} )}
\right) \Bigg]^{1/2}, ~~p \geq m+1 \nn
\end{align}

\begin{align}
N^p_\mu =  \Bigg[ &\prod_{k\neq \mu = 1}^m \left( \frac
{(\alpha_{k,p} - \alpha_{\mu,p} +1)(\alpha_{k,p} - \alpha_{\mu,p} + 2)}
{(\alpha_{k,p-1} - \alpha_{\mu,p} + 1)(\alpha_{k,p+1} - \alpha_{\mu,p} + 2)}
\right) \nn\\
\times 
&\left( \frac
{ \prod^{p-m-1}_{\nu = 1} (\alpha_{\nu,p-1} - \alpha_{\mu,p} ) \prod_{\nu=1}^{p-m+1} (\alpha_{\nu,p+1} - \alpha_{\mu,p} + 1)}
{ \prod^{p-m}_{\nu \neq \mu = 1} (\alpha_{\nu,p} - \alpha_{\mu,p}  )({\alpha_{\nu,p} - \alpha_{\mu,p} + 1} )}
\right) \Bigg]^{1/2}, ~~p \geq m+1. \nn
\end{align}

Note that for the case $p=m$ we have
\begin{align}
N^m_r ( \lambda_{q,m+1} ; \lambda_{q,m} ; \lambda_{q,m-1}) = 
\langle \lambda_{q,s} | \bar{\delta}_{r,m} \bar{c}_{r,m} |  \lambda_{q,s} \rangle^{1/2} \nn
\end{align}
where $\bar{\delta}_{r,m}$ is a $gl(m|1)$ invariant but $\bar{c}_{r,m}$ is a $gl(m)$
invariant. That is, $\bar{c}_{r,m}$ is dependent only on labels at rows $m$ and $m-1$ of
the GT pattern. These two rows of labels satisfy the usual $gl(m)$ branching conditions.
We may therefore obtain $\bar{c}_{r,m}$ from \cite{GIW1} by allowing the index sets $I$ and $I'$
to range over all possible values and setting all parity factors to be $0$ (even) as shown
below
\begin{align}
\bar{c}_r &= \prod_{k\in \tilde{I}',k\neq r} \left(\bar{\beta}_r - \bar{\beta}_k\right)^{-1}\prod_{k\in
I'} \left(\bar{\beta}_r - \bar{\alpha}_k - (-1)^{(k)}\right),\ \ r\in \tilde{I}' \nn\\
\Rightarrow
  	\bar{c}_{i,p} &= \prod^{p}_{k \neq i} (\bar{\alpha}_{i,p} -
\bar{\alpha}_{k,p})^{-1} \prod_{k=1}^{p-1} (\bar{\alpha}_{i,p} - \bar{\alpha}_{k,p-1} - 1),~i \leq p \leq m \nn\\
  	&= \prod^{p}_{k \neq i} (\alpha_{k,p} - \alpha_{i,p} )^{-1} \prod_{k=1}^{p-1} (\alpha_{k,p-1} - \alpha_{i,p} ),~i \leq p \leq m. \nn
\end{align}
The above formula is consistent with that previously obtained for the $U(m)$ case in \cite{Gould1981}. 

Continuing with the $p=m$ case we are able to utilize the $\bar{\delta}_{i}$ equations given earlier but with $n=0$ 
\begin{align}
\bar{\delta}_{i,m} &= (\beta_{m+1}-\alpha_i-1) \prod_{k\neq
i}^m\left(\frac{\alpha_k-\alpha_i}{\beta_k-\alpha_i}  \right)
, \ \ 1\leq i\leq m.\nn
\end{align}
After the change in notation $\beta_a \rightarrow \alpha_{a,m+1}$ and $\alpha_a \rightarrow \alpha_{a,m}$ we have
\begin{align}
\bar{\delta}_{i,m} &= (\alpha_{m+1,m+1}-\alpha_{i,m}-1) \prod_{k\neq
i}^m\left(\frac{\alpha_{k,m}-\alpha_{i,m}}{\alpha_{k,m+1}-\alpha_{i,m}}  \right)
, \ \ 1\leq i\leq m,\nn
\end{align}
allowing us to give the squared matrix element as
\begin{align}
\left( N^m_i \right)^2 &= \bar{\delta}_{i,m} \bar{c}_{i,m}  \nn\\
&=  (\alpha_{m+1,m+1}-\alpha_{i,m}-1) \prod_{k\neq
i}^m\left(\frac{\alpha_{k,m}-\alpha_{i,m}}{\alpha_{k,m+1}-\alpha_{i,m}}  \right) \prod^{m}_{k \neq i} (\alpha_{k,m} - \alpha_{i,m})^{-1} \prod_{k}^{m-1} (\alpha_{k,m-1} - \alpha_{i,m}) \nn \\
&=  (\alpha_{m+1,m+1}-\alpha_{i,m}-1) \frac{\prod_{k}^{m-1} (\alpha_{k,m-1} - \alpha_{i,m})}{ \prod_{k\neq
i}^m\left(\alpha_{k,m+1}-\alpha_{i,m} \right) }. \nn 
\end{align}
The final matrix element formula is then
\begin{align}
N^m_i =   (\alpha_{m+1,m+1}-\alpha_{i,m}-1)^{1/2} \left( \frac{\prod_{k}^{m-1} (\alpha_{k,m-1} - \alpha_{i,m})}{ \prod_{k\neq
i}^m\left(\alpha_{k,m+1}-\alpha_{i,m} \right) } \right)^{1/2}. \nn 
\end{align}

For $p < m$ we can utilize the results for the $U(m)$ case in \cite{Gould1981}. This matrix element formula is given here for convenience:
\begin{align}
	N^p_r =  \left( \frac{(-1)^p \prod_{k = 1}^{p+1} (\alpha_{k,p+1} - \alpha_{r,p} -
1) \prod_{k=1}^{p-1} (\alpha_{r,p} - \alpha_{k,p-1})}{\prod_{k \neq r}^p (\alpha_{r,p} -
\alpha_{k,p} + 1) (\alpha_{r,p} - \alpha_{k,p})  } \right)^{1/2} , \ \ p < m.\nn
\end{align}

The method of obtaining the matrix elements of the non-elementary generators $E_{l,p+1}$ is similar. 
Resolving $E_{l,p+1}$ $(l < p)$ into simultaneous shift components, we have
\begin{align}
E_{l,p+1} | \lambda_{q,s} \rangle &= \sum_u \psi[u_{p}u_{p-1}\ldots u_{l+1}u_l]^l | \lambda_{q,s} \rangle \nn\\
&= \sum_u N[u_{p},u_{p-1},\ldots ,u_{l+1},u_l]  
| \lambda_{q,s} + \Delta_{u_p,p} + \ldots + \Delta_{u_l,l} \rangle, \nn
\end{align}
where $| \lambda_{q,s} + \Delta_{u_p,p} + \ldots + \Delta_{u_l,l}  \rangle$ indicates the GT pattern 
obtained from $| \lambda_{q,s} \rangle$ by increasing the $p-l+1$ labels $\lambda_{u_r,r}$ of the 
subalgebra $gl(m|r-m)$ for $r=l,\ldots,p$, by one unit and leaving the remaining labels unchanged. 
Here the matrix elements
\begin{align}
N[u_{p},u_{p-1},\ldots, u_{l+1},u_l] \nn
\end{align}
are given by
\begin{align}
\pm \langle \lambda_{q,s} | \psi^\dagger [u_{p}u_{p-1}\ldots u_{l+1}u_l]_l 
\psi [u_{p}u_{p-1}\ldots u_{l+1}u_l]^l | \lambda_{q,s} \rangle ^ {1/2}.   \nn
\end{align}
Therefore, from equations (\ref{equ6.4}) and (\ref{equ6.7}) , we can express this matrix element as
\begin{align}
N[u_{p},u_{p-1},\ldots, u_l] 
&= 
\pm \left( \bar{\delta}_{u_p}\bar{c}_{u_l} \right)^{1/2} 
\prod_{s=l+1}^p \left(\bar{\rho}_{u_s,u_{s-1}}\right)^{1/2} 
\nn\\
&= \pm \left( \bar{\delta}_{u_p}\bar{c}_{u_l} \right)^{1/2} 
\prod_{s=l+1}^p \left[-(\bar{\beta}_{u_s}-\bar{\alpha}_{u_{s-1}} + 1)^{-1}
(\bar{\beta}_{u_s}-\bar{\alpha}_{u_{s-1}})^{-1} \bar{c}_{u_s}  \bar{\delta}_{u_{s-1}} \right]^{1/2}  
\label{NonSimpleRaising}\\
&= \pm \prod_{r=l}^p N^r_{u_r}  
\prod_{s=l+1}^p   \left[(\bar{\beta}_{u_s}-\bar{\alpha}_{u_{s-1}} + 1)^{-1}
(\bar{\beta}_{u_s}-\bar{\alpha}_{u_{s-1}})^{-1} \right]^{1/2}  
\label{NonSimpleRaising2}
\end{align}
where the undefined sign is specified later in Section \ref{phases} and we have
\begin{align}
\bar{\beta}_{u_s} - \bar{\alpha}_{u_{s-1}} 
= (-1)^{(u_{s-1})} (\lambda_{u_{s-1}} - u_{s-1}) - (-1)^{(u_{s})}(\lambda_{u_{s}} - u_s) +
((-1)^{(u_{s-1})} - (-1)^{(u_{s})})(m+1) 
\label{rootdiff} 
\end{align}
since
\begin{align}
&\bar{\beta}_r = (-1)^{(r)}(-\lambda_r + r-m-1) + m, \nn\\
&\bar{\alpha}_k = (-1)^{(k)}(-\lambda_k + k-m-1) + m. \nn
\end{align}
Also note within equation (\ref{NonSimpleRaising}), that the term 
$(\bar{\beta}_{u_s}-\bar{\alpha}_{u_{s-1}} -(-1)^{(u_{s-1})})^{-1}$ cancels the 
corresponding term within the numerator of $\bar{c}_{u_s}$.

We may now obtain matrix elements of the \textit{lowering} operators $E_{p+1,p}$ via the relation
\begin{align}
\langle (\lambda) - \delta_{rp}| E_{p+1,p} |(\lambda)\rangle  
= \overline{\langle (\lambda) | E_{p,p+1} |(\lambda - \delta_{rp})\rangle } \nn,
\end{align}
which holds on type 1 unitary representations.
We define the \textit{translated raising operator} $E_{p,p+1}'$ as
\begin{align}
\langle (\lambda + \delta_{rp}) | E_{p,p+1}' |(\lambda) \rangle  
& = \langle (\lambda) | E_{p,p+1} |(\lambda - \delta_{rp})\rangle  \nn
\end{align}
Since our matrix elements are real, the translated raising operator $E_{p,p+1}'$ is precisely 
the lowering operator $E_{p+1,p}$ we seek. It is clear that, $E_{p,p+1}'$ is simply obtained 
from  $E_{p,p+1}$ by making the substitution $\lambda_{rp} \rightarrow \lambda_{rp} - 1$ within the 
characteristic roots occuring in the matrix element formula for $E_{p,p+1}$. The final
result is presented below in Section \ref{resultsum}.

We now consider the matrix element $\bar{N}[u_{p},u_{p-1},\ldots, u_{l+1},u_l]$ of the
non-elementary lowering operators $E_{p+1,l}$ for $(l < p)$. 
The calculation is analogous to the $E_{l,p+1}$ case given above. 
We obtain
\begin{align}
\bar{N} [u_{p}, u_{p-1}, \ldots, u_l] &= \pm \left( \delta_{u_p}c_{u_l} \right)^{1/2} 
\prod_{s=l+1}^p \left(\rho_{u_s,u_{s-1}}\right)^{1/2} \nn\\
&= \pm \left( \delta_{u_p}c_{u_l} \right)^{1/2} \prod_{s=l+1}^p \left[-(\beta_{u_s}-\alpha_{u_{s-1}} + 1)^{-1}
(\beta_{u_s}-\alpha_{u_{s-1}})^{-1} c_{u_s}\delta_{u_{s-1}} \right]^{1/2}  \nn\\
&= \pm \prod_{r=l}^p \bar{N}^r_{u_r}  \prod_{s=l+1}^p  \left[(\beta_{u_s}-\alpha_{u_{s-1}} + 1)^{-1}
(\beta_{u_s}-\alpha_{u_{s-1}})^{-1} \right]^{1/2},  \label{Nbarform}
\end{align}
where
\begin{align}
\beta_{u_s} - \alpha_{u_{s-1}}  = (-1)^{(u_{s})}(\lambda_{u_{s}} - u_s) - (-1)^{(u_{s-1})}
(\lambda_{u_{s-1}} - u_{s-1}) - ((-1)^{(u_{s-1})} - (-1)^{(u_{s})})m - 1. 
\label{rootdifftwo} 
\end{align}

{\bf Remarks: } \label{importantremarks}
\begin{enumerate}
\item In view of the remark on page \pageref{remarkref},we may observe that in
in the formula (\ref{NonSimpleRaising2}) for the non-elementary raising generators, making the
substitutions $N^r_{u_r}$ $\rightarrow$ $\bar{N}^r_{u_r}$, $\bar{\beta}_{u_s}$
$\rightarrow$ $\beta_{u_s}$,
$\bar{\alpha}_{u_{s-1}}$ $\rightarrow$ $\alpha_{u_{s-1}}$ clearly gives the matrix element
formula (\ref{Nbarform}) for the non-elementary lowering generators. This further
highlights the symmetry of the expressions for the matrix elements in interchange of the two 
types of characteristic roots $\bar{\alpha}_r$ and $\alpha_r$.
\item
It is understood that to apply the matrix element formula derived above, where possible
terms are cancelled first and reduced to the most simplified rational form before
applying the formulae and substituting weight labels.
\item
All terms appearing in the square roots in the above formula are indeed positive numbers.
\item
We remind the reader that in all cases we have adopted the convention that a shifted
pattern $| \lambda_{q,s}' \rangle$ is identically zero if the branching rules are not satisfied. In
particular, the matrix element corresponding to a forbidden GT pattern (i.e. one for which
the branching rules of Theorem \ref{mainbranchingrule} are not satisfied) is identically
zero.
\end{enumerate}

\subsection{Phases} \label{phases}

The as yet undetermined sign in the above matrix element equations will now be examined. 
Following the Baird and Biedenharn phase convention, we choose the phases of the 
generators $E_{p,p+1}$ to be real and positive. By Hermiticity, the phases of the
generators $E_{p+1,p}$ are also 
positive. The phases of the remaining non-elementary generators may then be calculated via the algebra commutation 
relations. The phases of the matrix elements
\begin{align}
N[u_{p},u_{p-1},\ldots, u_{l+1},u_l],~\bar{N}[u_{p},u_{p-1}\ldots, u_{l+1},u_l] \nn
\end{align}
are then given by the expression
\begin{align}
S(\bar{N}[u_p,u_{p-1},\ldots,u_l])=S(N[u_p,u_{p-1},\ldots,u_l])\equiv\prod^p_{s=l+1} (-1)^{(u_{s-1})(u_s)} S(u_s - u_{s-1})
\label{phase} 
\end{align}
where $S(x) \in \{-1,1\}$ is the sign of $x$, $S(0) = 1$ and, as usual, odd indices are
considered greater than even indices. The details can be found in Appendix B.

\section{Summary of main results} \label{resultsum}

Here we summarise the main results that have been derived in the current section, by
presenting the matrix element formulae for the generators of $gl(m|n+1)$ in an irreducible
type 1 unitary representation. The basis vectors can be expressed in the form
(\ref{genGT}), and we now give expressions in terms of the weight labels $\lambda_{q,s}$,
using the characteristic root equations (\ref{evenalphabar}), (\ref{oddalphabar}),
(\ref{evenalpha}) and (\ref{oddalpha}). Note also that the labels $\lambda_{q,s}$
determining the basis vectors are subject to the branching rules of Theorem
\ref{mainbranchingrule}.

For generators $E_{p+1,p+1},$ $1\leq p\leq m+n$, the matrices are diagonal with
$$
\sum_{r=1}^{p+1} \lambda_{r,p+1} - \sum_{r=1}^{p} \lambda_{r,p}
$$
as the entry coinciding with the vector $\ket{\lambda_{q,s}}$ in the ordered basis.
Similarly, the matrix of the generator $E_{11}$ has entries $\lambda_{11}$ on the diagonal.

For raising generators, our derivation makes use of the characteristic matrix $\bar{{\cal
A}}$ of equation (\ref{adjointA}). The matrix elements of the elementary raising
generators $E_{p,p+1}$, $1\leq p\leq m+n$ are determined by
$$
E_{p,p+1} | \lambda_{q,s} \rangle 
= \sum_{r=1}^p N^p_r ( \lambda_{q,p+1} ; \lambda_{q,p} ; \lambda_{q,p-1}) 
| \lambda_{q,s} + \Delta_{r,p} \rangle, \nn
$$
with $N^p_r$ given in terms of the GT basis labels as follows.
For $p\geq m+1$,
\begin{align}
N^p_i &=  \Bigg[ \prod_{k\neq i = 1}^m \left( \frac
{(\lambda_{k,p} - \lambda_{i,p} - k + i - 1 )(\lambda_{k,p} - \lambda_{i,p} - k + i  )}
{(\lambda_{k,p-1} - \lambda_{i,p} - k + i )(\lambda_{k,p+1} - \lambda_{i,p} - k + i - 1 )}
\right) \nn\\
& \quad\times \left( \frac
{ \prod^{p-m-1}_{\nu = 1} (-\lambda_{\nu,p-1} - \lambda_{i,p} + \nu + i - m - 1)  \prod_{\nu=1}^{p-m+1} (-\lambda_{\nu,p-m+1} - \lambda_{i,p} + \nu + i - m - 2)}
{ \prod^{p-m}_{\nu = 1} (-\lambda_{\nu,p} - \lambda_{i,p} + \nu + i - m - 2 )(-\lambda_{\nu,p} - \lambda_{i,p} + \nu + i - m - 1 )}
\right) \Bigg]^{1/2}, \nn
\end{align}
\begin{align}
N^p_\mu &=\Bigg[ \prod_{k = 1}^m \left( \frac
{(\lambda_{k,p} + \lambda_{\mu,p} - k - \mu + m + 1)( \lambda_{k,p} + \lambda_{\mu,p} - k - \mu + m + 2 )}
{(\lambda_{k,p-1} +\lambda_{\mu,p} - k  - \mu + m + 2)(\lambda_{k,p+1} + \lambda_{\mu,p} - k  - \mu + m + 1  )}
\right) \nn\\
& \quad \times \left( \frac
{ \prod^{p-m-1}_{\nu = 1} (-\lambda_{\nu,p-1} +\lambda_{\mu,p} + \nu  - \mu + 1)  \prod_{\nu=1}^{p-m+1} ( -\lambda_{\nu,p+1} + \lambda_{\mu,p} + \nu  - \mu )}
{ \prod^{p-m}_{\nu \neq \mu = 1} (-\lambda_{\nu,p} + \lambda_{\mu,p} + \nu - \mu )(-\lambda_{\nu,p} + \lambda_{\mu,p} + \nu - \mu + 1 )}
\right) \Bigg]^{1/2}. \nn
\end{align}
We also have (for the case $p=m$)
\begin{align}
N^m_i =   (-\lambda_{m+1,m+1}-\lambda_{i,m} - m  + i -1 )^{1/2} \left( \frac{\prod_{k}^{m-1} (\lambda_{k,m-1} - \lambda_{i,m} - k + i-1)}{ \prod_{k\neq
i}^m\left(\lambda_{k,m+1}-\lambda_{i,m} - k + i -1 \right) } \right)^{1/2}. \nn 
\end{align}
Finally, for $p<m$,
\begin{align}
	N^p_i =  \left( \frac{(-1)^p \prod_{k = 1}^{p+1} (\lambda_{k,p+1} - \lambda_{i,p}
- k + i) \prod_{k=1}^{p-1} (\lambda_{i,p} - \lambda_{k,p-1} - k + i + 1)}{\prod_{k \neq
i}^p (\lambda_{i,p} - \lambda_{k,p} + k - i + 1) (\lambda_{i,p} - \lambda_{k,p} + k - i)
} \right)^{1/2},\nn
\end{align}
which was derived for the Lie algebra case in \cite{Gould1981}.
The above matrix element equations are valid for all type 1 unitary 
irreducible representations. Our matrix element equations match those given by Palev \cite{Palev1989} 
and Stoilova and Van der Jeugt \cite{StoiVan2010} where they each considered a subclass of
these representations. As mentioned previously, we have adopted the convention that the
phase of the matrix elements of the elementary generators are real and positive.

We may also give explicit expressions for the non-elementary raising generators
$E_{l,p+1}$, $l<p$, with
$$
E_{l,p+1} | \lambda_{q,s} \rangle = \sum_u N[u_{p},u_{p-1},\ldots, u_{l+1},u_l] 
| \lambda_{q,s} + \Delta_{u_p,p} + \ldots + \Delta_{u_l,l} \rangle, 
$$
where the sum is over all allowable shift components, as we have already described in
equations (\ref{equ6.5a}) and (\ref{equ6.5b}). 
In this case the matrix elements take on the form
$$
N[u_{p},u_{p-1},\ldots, u_l] 
= \frac{S(N[u_p,u_{p-1},\ldots,u_l]) \prod_{r=l}^p N^r_{u_r}}{
\prod_{s=l+1}^p  \sqrt{(\bar{\beta}_{u_s}-\bar{\alpha}_{u_{s-1}} + 1)
(\bar{\beta}_{u_s}-\bar{\alpha}_{u_{s-1}}) } },
$$
where the difference $\bar{\beta}_{u_s} - \bar{\alpha}_{u_{s-1}} $ in characteristic roots
has been given in terms of the labels $\lambda_{q,s}$ in equation (\ref{rootdiff}), and
the phase $S(N[u_p,u_{p-1},\ldots,u_l])$ is given in equation (\ref{phase}). 

For the lowering generators, we first give the elementary generators $E_{p+1,p}$,
$1\leq p\leq m+n$, with
$$
E_{p+1,p} | \lambda_{q,s} \rangle 
= \sum_{r=1}^p \bar{N}^p_r ( \lambda_{q,p+1} ; \lambda_{q,p} ; \lambda_{q,p-1}) 
| \lambda_{q,s} - \Delta_{r,p} \rangle. 
$$
For $p\geq m+1$, 
\begin{align}
\bar{N}^p_i =  \Bigg[ &\prod_{k\neq i = 1}^m \left( \frac
{(\lambda_{k,p} - \lambda_{i,p} - k + i  )(\lambda_{k,p} - \lambda_{i,p} - k + i + 1 )}
{(\lambda_{k,p-1} - \lambda_{i,p} - k + i + 1 )(\lambda_{k,p+1} - \lambda_{i,p} - k + i )}
\right) \nn\\
\times &\left( \frac
{ \prod^{p-m-1}_{\nu = 1} (-\lambda_{\nu,p-1} - \lambda_{i,p} + \nu + i - m )  
\prod_{\nu=1}^{p-m+1} (-\lambda_{\nu,p+1} - \lambda_{i,p} + \nu + i - m - 1)}
{ \prod^{p-m}_{\nu = 1} (-\lambda_{\nu,p} - \lambda_{i,p} + \nu + i - m - 1 )
(-\lambda_{\nu,p} - \lambda_{i,p} + \nu + i - m  )}
\right) \Bigg]^{1/2}, \nn
\end{align}
\begin{align}
\bar{N}^p_\mu =\Bigg[ &\prod_{k = 1}^m \left( \frac
{(\lambda_{k,p} + \lambda_{\mu,p} - k - \mu + m )( \lambda_{k,p} + \lambda_{\mu,p} - k - \mu + m + 1)}
{(\lambda_{k,p-1} +\lambda_{\mu,p} - k  - \mu + m + 1)(\lambda_{k,p+1} + \lambda_{\mu,p} - k  - \mu + m  )}
\right) \nn\\
\times &\left( \frac
{\prod^{p-m-1}_{\nu = 1} (-\lambda_{\nu,p-1} +\lambda_{\mu,p} + \nu  - \mu )  
\prod_{\nu=1}^{p-m+1} ( -\lambda_{\nu,p+1} + \lambda_{\mu,p} + \nu  - \mu - 1 )}
{ \prod^{p-m}_{\nu \neq \mu = 1} (-\lambda_{\nu,p} + \lambda_{\mu,p} + \nu - \mu - 1 )
(-\lambda_{\nu,p} + \lambda_{\mu,p} + \nu - \mu  )} 
\right) \Bigg]^{1/2}. \nn
\end{align}
For the case $p=m$ we have
\begin{align}
\bar{N}^m_i =  (-\lambda_{m+1,m+1}-\lambda_{i,m} - m  + i )^{1/2} \left( \frac{\prod_{k}^{m-1} (\lambda_{k,m-1} - \lambda_{i,m} - k + i )}{ \prod_{k\neq
i}^m\left(\lambda_{k,m+1}-\lambda_{i,m} - k + i  \right) } \right)^{1/2}. \nn 
\end{align}
Finally, when $p<m$, we once again make use of the results in \cite{Gould1985}, namely
\begin{align}
	\bar{N}^p_i =  \left( \frac{(-1)^p \prod_{k = 1}^{p+1} (\lambda_{k,p+1} -
\lambda_{i,p} - k + i + 1) \prod_{k=1}^{p-1} (\lambda_{i,p} - \lambda_{k,p-1} - k + i
)}{\prod_{k \neq i}^p (\lambda_{i,p} - \lambda_{k,p} + k - i ) (\lambda_{i,p} -
\lambda_{k,p} + k - i - 1)  } \right)^{1/2}.\nn
\end{align}

The matrix elements $\bar{N}[u_{p},u_{p-1},\ldots, u_{l+1},u_l]$ of the
non-elementary lowering operators $E_{p+1,l}$ for $(l < p)$ appears as
$$
E_{p+1,l} | \lambda_{q,s} \rangle = \sum_u \bar{N}[u_{p},u_{p-1},\ldots, u_{l+1},u_l] 
| \lambda_{q,s} - \Delta_{u_p,p} - \ldots - \Delta_{u_l,l} \rangle,
$$
and are given by 
\begin{align}
\bar{N} [u_{p}, u_{p-1} ,\ldots, u_l] 
=\frac{S(\bar{N}[u_p,u_{p-1},\ldots,u_l]) \prod_{r=l}^p \bar{N}^r_{u_r}}{
\prod_{s=l+1}^p  \sqrt{(\beta_{u_s}-\alpha_{u_{s-1}} + 1)
(\beta_{u_s}-\alpha_{u_{s-1}}) } },
\nn\end{align}
where the difference $\beta_{u_s} - \alpha_{u_{s-1}}$ in characteristic roots is given by
equation (\ref{rootdifftwo}), and the phase $S(\bar{N}[u_p,u_{p-1},\ldots,u_l])$ is once again given in equation (\ref{phase}).  

Once again, we remind the reader of the remarks beginning on page
\pageref{importantremarks}. In particular, any pattern not satisfying the branching rules
of Theorem \ref{mainbranchingrule} after a shift in labels is identically zero, and hence
so is the corresponding matrix element. In the above results, the patterns to which this
comment pertains are of the form 
$$| \lambda_{q,s} + \Delta_{r,p} \rangle,\ \  
|\lambda_{q,s} + \Delta_{u_p,p} + \ldots + \Delta_{u_l,l} \rangle,\ \ 
| \lambda_{q,s} - \Delta_{r,p} \rangle \mbox{ and } |\lambda_{q,s} - \Delta_{u_p,p} -
\ldots - \Delta_{u_l,l} \rangle.
$$


\section{Example: Matrix elements of the $gl(2|2)$ raising generators} \label{example}

As an explicit example, we calculate the action (and hence matrix elements) of each
raising generator of $gl(2|2)$ on a GT basis vector in an arbitrary irreducible type 1
unitary module with highest weight $(\lambda_{14},\lambda_{24}\ |\
\lambda_{\bar{1}4},\lambda_{\bar{2}4})$.

In what follows, we remind the reader of the convention in place that if a vector with shifted labels is no
longer a genuine GT pattern satisfying the branching rules of Theorem
\ref{mainbranchingrule}, then it is identically zero, regardless of how the expressions
for the coefficients turn out. Hence the corresponding matrix element would also be
identically zero in such a case.

\subsection{Elementary generators $E_{i,i+1}$}
\begin{align}
E_{1,2} \left\lvert
\begin{matrix}
 \lambda_{1,4}&\lambda_{2,4}&\lambda_{\bar{1},4}&\lambda_{\bar{2},4}& \\
 \lambda_{1,3}&\lambda_{2,3}&\lambda_{\bar{1},3}& & \\
 \lambda_{1,2}&\lambda_{2,2}& & & \\ 
 \lambda_{1,1}& & & & \\
\end{matrix}
\right) 
&= N^1_1 \left\lvert
\begin{matrix}
 \lambda_{1,4}&\lambda_{2,4}&\lambda_{\bar{1},4}&\lambda_{\bar{2},4}& \\
 \lambda_{1,3}&\lambda_{2,3}&\lambda_{\bar{1},3}& & \\
 \lambda_{1,2}&\lambda_{2,2}& & & \\ 
 \lambda_{1,1} + 1& & & & \\
\end{matrix}
\right) \nn
\end{align}
where
\begin{align}
N^1_1 &= \sqrt{(\lambda_{1,2} - \lambda_{1,1} )(\lambda_{1,1} - \lambda_{2,2} + 1)} . \nn
\end{align}

\begin{align}
E_{2,3} \left\lvert
\begin{matrix}
 \lambda_{1,4}&\lambda_{2,4}&\lambda_{\bar{1},4}&\lambda_{\bar{2},4}& \\
 \lambda_{1,3}&\lambda_{2,3}&\lambda_{\bar{1},3}& & \\
 \lambda_{1,2}&\lambda_{2,2}& & & \\ 
 \lambda_{1,1}& & & & \\
\end{matrix}
\right) 
&= N^2_1 \left\lvert
\begin{matrix}
 \lambda_{1,4}&\lambda_{2,4}&\lambda_{\bar{1},4}&\lambda_{\bar{2},4}& \\
 \lambda_{1,3}&\lambda_{2,3}&\lambda_{\bar{1},3}& & \\
 \lambda_{1,2} + 1&\lambda_{2,2}& & & \\ 
 \lambda_{1,1}& & & & \\
\end{matrix}
\right) \nn\\
&~+ N^2_2 \left\lvert
\begin{matrix}
 \lambda_{1,4}&\lambda_{2,4}&\lambda_{\bar{1},4}&\lambda_{\bar{2},4}& \\
 \lambda_{1,3} &\lambda_{2,3}&\lambda_{\bar{1},3}& & \\
 \lambda_{1,2}&\lambda_{2,2} + 1& & & \\ 
 \lambda_{1,1}& & & & \\
\end{matrix}
\right) \nn
\end{align}
where
\begin{align}
N^2_1 &= \sqrt{\frac{ (\lambda_{\bar{1},3}+\lambda_{1,3} + 1 )(\lambda_{1,1} - \lambda_{1,3} )}{ \left(\lambda_{2,3}-\lambda_{1,3} - 1 \right) } },  \nn \\
N^2_2 &=  \sqrt{\frac{ (\lambda_{\bar{1},3}+\lambda_{2,3}  )(\lambda_{1,1} - \lambda_{2,3} + 1 )}{ \left(\lambda_{1,3}-\lambda_{2,3} + 1 \right) } }. \nn
\end{align}

\begin{align}
E_{3,4} \left\lvert
\begin{matrix}
 \lambda_{1,4}&\lambda_{2,4}&\lambda_{\bar{1},4}&\lambda_{\bar{2},4}& \\
 \lambda_{1,3}&\lambda_{2,3}&\lambda_{\bar{1},3}& & \\
 \lambda_{1,2}&\lambda_{2,2}& & & \\ 
 \lambda_{1,1}& & & & \\
\end{matrix}
\right) 
&= N^3_1 \left\lvert
\begin{matrix}
 \lambda_{1,4}&\lambda_{2,4}&\lambda_{\bar{1},4}&\lambda_{\bar{2},4}& \\
 \lambda_{1,3} + 1&\lambda_{2,3}&\lambda_{\bar{1},3}& & \\
 \lambda_{1,2}&\lambda_{2,2}& & & \\ 
 \lambda_{1,1}& & & & \\
\end{matrix}
\right) \nn\\
&~+ N^3_2 \left\lvert
\begin{matrix}
 \lambda_{1,4}&\lambda_{2,4}&\lambda_{\bar{1},4}&\lambda_{\bar{2},4}& \\
 \lambda_{1,3} &\lambda_{2,3} + 1 &\lambda_{\bar{1},3}& & \\
 \lambda_{1,2}&\lambda_{2,2}& & & \\ 
 \lambda_{1,1}& & & & \\
\end{matrix}
\right) \nn\\
&~+ N^3_3 \left\lvert
\begin{matrix}
 \lambda_{1,4}&\lambda_{2,4}&\lambda_{\bar{1},4}&\lambda_{\bar{2},4}& \\
 \lambda_{1,3} &\lambda_{2,3}&\lambda_{\bar{1},3} + 1 & & \\
 \lambda_{1,2}&\lambda_{2,2}& & & \\ 
 \lambda_{1,1}& & & & \\
\end{matrix}
\right) \nn
\end{align}
where
\begin{align}
N^3_1 &= \sqrt{ \frac{(\lambda_{2,3} - \lambda_{1,3} - 2)(\lambda_{2,3} - \lambda_{1,3} -
1)(\lambda_{\bar{1},4} + \lambda_{1,3} + 2)(\lambda_{\bar{2},4} + \lambda_{1,3} + 1)}
{(\lambda_{2,2} - \lambda_{1,3} - 1)(\lambda_{2,4} - \lambda_{1,3} -
2)(\lambda_{\bar{1},3} + \lambda_{1,3} + 2)(\lambda_{\bar{1},3} + \lambda_{1,3} + 1) } }, 
\nn\\
N^3_2 &= \sqrt{ \frac{(\lambda_{1,3} - \lambda_{2,3} )(\lambda_{1,3} - \lambda_{2,3} + 1)(-\lambda_{\bar{1},4} - \lambda_{2,3} - 1 )(-\lambda_{\bar{2},4} - \lambda_{2,3} )}
{(\lambda_{1,2} - \lambda_{2,3} + 1)(\lambda_{1,4} - \lambda_{2,3} )(-\lambda_{\bar{1},3} - \lambda_{2,3} - 1)(-\lambda_{\bar{1},3} - \lambda_{2,3} ) } }, \nn\\
N^3_3 &= \sqrt{ \frac{ (\lambda_{1,3} + \lambda_{\bar{1},3} + 1)(\lambda_{1,3} +
\lambda_{\bar{1},3} + 2)(\lambda_{2,3} + \lambda_{\bar{1},3} )(\lambda_{2,3} +
\lambda_{\bar{1},3} + 1)(\lambda_{\bar{1},4} - \lambda_{\bar{1},3})(\lambda_{\bar{2},4} - \lambda_{\bar{1},3} - 1)   }
{ (\lambda_{1,2} + \lambda_{\bar{1},3} + 2)(\lambda_{1,4} + \lambda_{\bar{1},3} +
1)(\lambda_{2,2} + \lambda_{\bar{1},3} + 1)(\lambda_{2,4} + \lambda_{\bar{1},3} ) }}. \nn
\end{align}

\subsection{Non-elementary generators}

\begin{align}
E_{1,3} \left\lvert
\begin{matrix}
 \lambda_{1,4}&\lambda_{2,4}&\lambda_{\bar{1},4}&\lambda_{\bar{2},4}& \\
 \lambda_{1,3}&\lambda_{2,3}&\lambda_{\bar{1},3}& & \\
 \lambda_{1,2}&\lambda_{2,2}& & & \\ 
 \lambda_{1,1}& & & & \\
\end{matrix}
\right) 
&= N^{2~1}_{1~1} \left\lvert
\begin{matrix}
 \lambda_{1,4}&\lambda_{2,4}&\lambda_{\bar{1},4}&\lambda_{\bar{2},4}& \\
 \lambda_{1,3}&\lambda_{2,3}&\lambda_{\bar{1},3}& & \\
 \lambda_{1,2} + 1&\lambda_{2,2}& & & \\ 
 \lambda_{1,1} + 1& & & & \\
\end{matrix}
\right) \nn\\ 
&+ N^{2~1}_{2~1} \left\lvert
\begin{matrix}
 \lambda_{1,4}&\lambda_{2,4}&\lambda_{\bar{1},4}&\lambda_{\bar{2},4}& \\
 \lambda_{1,3}&\lambda_{2,3}&\lambda_{\bar{1},3}& & \\
 \lambda_{1,2} &\lambda_{2,2} + 1& & & \\ 
 \lambda_{1,1} + 1& & & & \\
\end{matrix}
\right),
\nn
\end{align}
where
\begin{align}
N^{2~1}_{1~1} &= N^1_1 N^2_1 \sqrt{(\lambda_{1,1} - \lambda_{1,2} + 1)^{-1}(\lambda_{1,1} - \lambda_{1,2}  )^{-1}} \nn\\
&=  \sqrt{\frac{ (\lambda_{1,1} - \lambda_{2,2} + 1)(\lambda_{\bar{1},3}+\lambda_{1,3} + 1 )(\lambda_{1,1} - \lambda_{1,3} )}{  (\lambda_{2,3}-\lambda_{1,3} - 1 ) (\lambda_{1,1} - \lambda_{1,2} + 1) } } 
\nn 
\end{align}
and
\begin{align}
N^{2~1}_{2~1} &= -N^1_1 N^2_2 \sqrt{(\lambda_{1,1} - \lambda_{2,2} + 1)^{-1}( \lambda_{1,1} - \lambda_{2,2} )^{-1}} \nn\\
&= - \sqrt{\frac{ (\lambda_{1,2} - \lambda_{1,1} )(\lambda_{\bar{1},3}+\lambda_{2,3}
)(\lambda_{1,1} - \lambda_{2,3} + 1 )}{  \left(\lambda_{1,3}-\lambda_{2,3} + 1 \right)
(\lambda_{1,1} - \lambda_{2,2} ) } }.  \nn
\end{align}

\begin{align}
E_{2,4} \left\lvert
\begin{matrix}
 \lambda_{1,4}&\lambda_{2,4}&\lambda_{\bar{1},4}&\lambda_{\bar{2},4}& \\
 \lambda_{1,3}&\lambda_{2,3}&\lambda_{\bar{1},3}& & \\
 \lambda_{1,2}&\lambda_{2,2}& & & \\ 
 \lambda_{1,1}& & & & \\
\end{matrix}
\right) 
&= N^{3~2}_{1~1} \left\lvert
\begin{matrix}
 \lambda_{1,4}&\lambda_{2,4}&\lambda_{\bar{1},4}&\lambda_{\bar{2},4}& \\
 \lambda_{1,3} + 1&\lambda_{2,3}&\lambda_{\bar{1},3}& & \\
 \lambda_{1,2} + 1&\lambda_{2,2}& & & \\ 
 \lambda_{1,1}& & & & \\
\end{matrix}
\right) \nn\\ 
&+ N^{3~2}_{2~1} \left\lvert
\begin{matrix}
 \lambda_{1,4}&\lambda_{2,4}&\lambda_{\bar{1},4}&\lambda_{\bar{2},4}& \\
 \lambda_{1,3}&\lambda_{2,3}  + 1&\lambda_{\bar{1},3}& & \\
 \lambda_{1,2} + 1&\lambda_{2,2}& & & \\ 
 \lambda_{1,1}& & & & \\
\end{matrix}
\right) \nn\\ 
&+ N^{3~2}_{3~1} \left\lvert
\begin{matrix}
 \lambda_{1,4}&\lambda_{2,4}&\lambda_{\bar{1},4}&\lambda_{\bar{2},4}& \\
 \lambda_{1,3}&\lambda_{2,3}&\lambda_{\bar{1},3}  + 1& & \\
 \lambda_{1,2} + 1&\lambda_{2,2}& & & \\ 
 \lambda_{1,1}& & & & \\
\end{matrix}
\right) \nn\\
&+ N^{3~2}_{1~2} \left\lvert
\begin{matrix}
 \lambda_{1,4}&\lambda_{2,4}&\lambda_{\bar{1},4}&\lambda_{\bar{2},4}& \\
 \lambda_{1,3} + 1&\lambda_{2,3}&\lambda_{\bar{1},3}& & \\
 \lambda_{1,2}&\lambda_{2,2} + 1& & & \\ 
 \lambda_{1,1}& & & & \\
\end{matrix}
\right) \nn\\ 
&+ N^{3~2}_{2~2} \left\lvert
\begin{matrix}
 \lambda_{1,4}&\lambda_{2,4}&\lambda_{\bar{1},4}&\lambda_{\bar{2},4}& \\
 \lambda_{1,3}&\lambda_{2,3}  + 1&\lambda_{\bar{1},3}& & \\
 \lambda_{1,2}&\lambda_{2,2} + 1& & & \\ 
 \lambda_{1,1}& & & & \\
\end{matrix}
\right) \nn\\ 
&+ N^{3~2}_{3~2} \left\lvert
\begin{matrix}
 \lambda_{1,4}&\lambda_{2,4}&\lambda_{\bar{1},4}&\lambda_{\bar{2},4}& \\
 \lambda_{1,3}&\lambda_{2,3}&\lambda_{\bar{1},3}  + 1& & \\
 \lambda_{1,2}&\lambda_{2,2} + 1& & & \\
 \lambda_{1,1}& & & & \\
\end{matrix}
\right) \nn
\nn
\end{align}
where
\begin{align}
N^{3~2}_{1~1} &= N^2_1 N^3_1 \nn\\
&= \sqrt{\frac{ (\lambda_{1,1} - \lambda_{1,3} ) (\lambda_{2,3} -
\lambda_{1,3})(\lambda_{\bar{1},4} + \lambda_{1,3} + 1)(\lambda_{\bar{2},4} +
\lambda_{1,3}) }{(\lambda_{2,2} - \lambda_{1,3})(\lambda_{2,4} - \lambda_{1,3} -
1)(\lambda_{\bar{1},3} + \lambda_{1,3})  }}, \nn
\end{align}
\begin{align}
N^{3~2}_{2~1} &= N^2_1 N^3_2 \sqrt{(\lambda_{1,2} - \lambda_{2,3} + 2)^{-1}(\lambda_{1,2} - \lambda_{2,3} + 1 )^{-1}} \nn\\
&= \sqrt{\frac{ (\lambda_{\bar{1},3}+\lambda_{1,3} + 1 )(\lambda_{1,3} - \lambda_{1,1} )
(\lambda_{1,3} - \lambda_{2,3} )(\lambda_{\bar{1},4} + \lambda_{2,3} + 1
)(\lambda_{\bar{2},4} + \lambda_{2,3} )  }{ (\lambda_{1,2} - \lambda_{2,3} +
1)(\lambda_{1,4} - \lambda_{2,3} )(\lambda_{\bar{1},3} + \lambda_{2,3} +
1)(\lambda_{\bar{1},3} + \lambda_{2,3} )(\lambda_{1,2} - \lambda_{2,3} + 2)(\lambda_{1,2}
- \lambda_{2,3} + 1) } }, \nn
\end{align}
\begin{align}
N^{3~2}_{3~1} &= N^2_1 N^3_3 \sqrt{(\lambda_{1,2} - \lambda_{\bar{1},3} + 3)^{-1}(\lambda_{1,2} - \lambda_{\bar{1},3} + 2 )^{-1}} 
\nn\\
&= \sqrt{\frac{ (\lambda_{\bar{1},3}+\lambda_{1,3} + 1 )^2(\lambda_{1,1} - \lambda_{1,3} )
                (\lambda_{1,3} + \lambda_{\bar{1},3} + 2)(\lambda_{2,3} + \lambda_{\bar{1},3} )(\lambda_{2,3} + \lambda_{\bar{1},3} + 1) }
{\left(\lambda_{2,3}-\lambda_{1,3} - 1 \right) (\lambda_{1,2} + \lambda_{\bar{1},3} + 2)(\lambda_{1,4} + \lambda_{\bar{1},3} + 1)(\lambda_{2,2} + \lambda_{\bar{1},3} + 1)(\lambda_{2,4} + \lambda_{\bar{1},3} ) } } 
\nn\\
& \qquad \times \sqrt{ \frac{(-\lambda_{\bar{1},4} + \lambda_{\bar{1},3})(-\lambda_{\bar{2},4} +
\lambda_{\bar{1},3} + 1)}
{(\lambda_{1,2} - \lambda_{\bar{1},3} + 3)(\lambda_{1,2} - \lambda_{\bar{1},3} + 2 )}},
\nn \\
N^{3~2}_{1~2} &= -N^2_2 N^3_1 \sqrt{(\lambda_{2,2} - \lambda_{1,3} )^{-1}(\lambda_{2,2} - \lambda_{1,3} - 1 )^{-1}}\nn\\
&= - \sqrt{\frac{ (\lambda_{\bar{1},3}+\lambda_{2,3}  )(\lambda_{1,1} - \lambda_{2,3} + 1
) (\lambda_{1,3} - \lambda_{2,3})(\lambda_{\bar{1},4} + \lambda_{1,3} +
1)(\lambda_{\bar{2},4} + \lambda_{1,3}) }
{ (\lambda_{2,2} - \lambda_{1,3})^2(\lambda_{2,4} -
\lambda_{1,3} - 1)(\lambda_{\bar{1},3} + \lambda_{1,3} + 1)(\lambda_{\bar{1},3} +
\lambda_{1,3})(\lambda_{2,2} - \lambda_{1,3} - 1 ) } }, \nn\\
N^{3~2}_{2~2} &= N^2_2 N^3_2  \nn\\
&= \sqrt{\frac{ (\lambda_{1,1} - \lambda_{2,3} + 1 ) (\lambda_{1,3} - \lambda_{2,3}
)(\lambda_{\bar{1},4} + \lambda_{2,3} + 1 )(\lambda_{\bar{2},4} + \lambda_{2,3} )  }
{(\lambda_{1,2} - \lambda_{2,3} + 1)(\lambda_{1,4} - \lambda_{2,3} )(\lambda_{\bar{1},3} + \lambda_{2,3} + 1) } } \nn
\end{align}
and
\begin{align}
N^{3~2}_{3~2} &= N^2_2 N^3_3 \sqrt{(\lambda_{2,2} - \lambda_{\bar{1},3} + 2)^{-1}(\lambda_{2,2} - \lambda_{\bar{1},3} + 1 )^{-1}} \nn\\
&=  \sqrt{\frac{ (\lambda_{\bar{1},3}+\lambda_{2,3}  )^2(\lambda_{1,1} - \lambda_{2,3} + 1
) (\lambda_{1,3} + \lambda_{\bar{1},3} + 1)(\lambda_{1,3} + \lambda_{\bar{1},3} + 2)(\lambda_{2,3} + \lambda_{\bar{1},3} + 1)  }
{ \left(\lambda_{1,3}-\lambda_{2,3} + 1 \right) (\lambda_{1,2} + \lambda_{\bar{1},3} + 2)(\lambda_{1,4} + \lambda_{\bar{1},3} +
1)(\lambda_{2,2} + \lambda_{\bar{1},3} + 1)(\lambda_{2,4} + \lambda_{\bar{1},3} )  } } \nn\\
& \qquad \times \sqrt{\frac{(\lambda_{\bar{1},4} -
\lambda_{\bar{1},3})(\lambda_{\bar{2},4} - \lambda_{\bar{1},3} - 1)}{ 
(\lambda_{2,2} - \lambda_{\bar{1},3} + 2)(\lambda_{2,2} - \lambda_{\bar{1},3} + 1 )}}. \nn
\end{align}
\begin{align}
E_{1,4} \left\lvert
\begin{matrix}
 \lambda_{1,4}&\lambda_{2,4}&\lambda_{\bar{1},4}&\lambda_{\bar{2},4}& \\
 \lambda_{1,3}&\lambda_{2,3}&\lambda_{\bar{1},3}& & \\
 \lambda_{1,2}&\lambda_{2,2}& & & \\ 
 \lambda_{1,1}& & & & \\
\end{matrix}
\right) 
&= N^{3~2~1}_{1~1~1} \left\lvert
\begin{matrix}
 \lambda_{1,4}&\lambda_{2,4}&\lambda_{\bar{1},4}&\lambda_{\bar{2},4}& \\
 \lambda_{1,3} + 1&\lambda_{2,3}&\lambda_{\bar{1},3}& & \\
 \lambda_{1,2} + 1&\lambda_{2,2}& & & \\ 
 \lambda_{1,1} + 1& & & & \\
\end{matrix}
\right) \nn\\ 
&+ 
N^{3~2~1}_{2~1~1} \left\lvert
\begin{matrix}
 \lambda_{1,4}&\lambda_{2,4}&\lambda_{\bar{1},4}&\lambda_{\bar{2},4}& \\
 \lambda_{1,3}&\lambda_{2,3}  + 1&\lambda_{\bar{1},3}& & \\
 \lambda_{1,2} + 1&\lambda_{2,2}& & & \\ 
 \lambda_{1,1} + 1& & & & \\
\end{matrix}
\right) \nn\\ 
&+ N^{3~2~1}_{3~1~1} \left\lvert
\begin{matrix}
 \lambda_{1,4}&\lambda_{2,4}&\lambda_{\bar{1},4}&\lambda_{\bar{2},4}& \\
 \lambda_{1,3}&\lambda_{2,3}&\lambda_{\bar{1},3} + 1& & \\
 \lambda_{1,2} + 1&\lambda_{2,2}& & & \\ 
 \lambda_{1,1} + 1& & & & \\
\end{matrix}
\right) \nn\\
&+ N^{3~2~1}_{1~2~1} \left\lvert
\begin{matrix}
 \lambda_{1,4}&\lambda_{2,4}&\lambda_{\bar{1},4}&\lambda_{\bar{2},4}& \\
 \lambda_{1,3} + 1&\lambda_{2,3}&\lambda_{\bar{1},3}& & \\
 \lambda_{1,2}&\lambda_{2,2} + 1& & & \\ 
 \lambda_{1,1} + 1& & & & \\
\end{matrix}
\right) \nn\\ 
&+ N^{3~2~1}_{2~2~1} \left\lvert
\begin{matrix}
 \lambda_{1,4}&\lambda_{2,4}&\lambda_{\bar{1},4}&\lambda_{\bar{2},4}& \\
 \lambda_{1,3}&\lambda_{2,3}  + 1&\lambda_{\bar{1},3}& & \\
 \lambda_{1,2}&\lambda_{2,2} + 1& & & \\ 
 \lambda_{1,1} + 1& & & & \\
\end{matrix}
\right) \nn\\ 
&+ N^{3~2~1}_{3~2~1} \left\lvert
\begin{matrix}
 \lambda_{1,4}&\lambda_{2,4}&\lambda_{\bar{1},4}&\lambda_{\bar{2},4}& \\
 \lambda_{1,3}&\lambda_{2,3}&\lambda_{\bar{1},3} + 1& & \\
 \lambda_{1,2}&\lambda_{2,2} + 1& & & \\ 
 \lambda_{1,1} + 1& & & & \\
\end{matrix}
\right) \nn
\end{align}
where
\begin{align}
N^{3~2~1}_{1~1~1} &= N^1_1 N^2_1 N^3_1 \sqrt{(\lambda_{1,1} - \lambda_{1,2} - 1)^{-1}(\lambda_{1,1} - \lambda_{1,2})^{-1}} \nn \\
&= \sqrt{\frac{(\lambda_{1,1} - \lambda_{2,2} + 1)(\lambda_{1,1} - \lambda_{1,3} )(\lambda_{2,3} - \lambda_{1,3} - 2)(\lambda_{\bar{1},4} +
\lambda_{1,3} + 2) (\lambda_{\bar{2},4} + \lambda_{1,3} + 1)  }
{ (\lambda_{2,2} - \lambda_{1,3} - 1)(\lambda_{1,3} - \lambda_{2,4} + 2)(\lambda_{\bar{1},3} + \lambda_{1,3}
+ 2)(\lambda_{1,1} - \lambda_{1,2} - 1)   }}, \nn
\end{align}
\begin{align}
N^{3~2~1}_{2~1~1} &= N^1_1 N^2_1 N^3_2 \sqrt{(\lambda_{1,1} - \lambda_{1,2} - 1)^{-1}(\lambda_{1,1} - \lambda_{1,2})^{-1}} \sqrt{(\lambda_{1,2} - \lambda_{2,3} )^{-1}(\lambda_{1,2} - \lambda_{2,3} - 1)^{-1}}\nn \\
&= \sqrt{\frac{(\lambda_{1,1} - \lambda_{2,2} + 1)(\lambda_{\bar{1},3}+\lambda_{1,3} + 1
)(\lambda_{1,1} - \lambda_{1,3} ) (\lambda_{1,3} - \lambda_{2,3} )  }
{ (\lambda_{1,2} -\lambda_{2,3} + 1)(\lambda_{1,4} - \lambda_{2,3} )(\lambda_{\bar{1},3} + \lambda_{2,3} +
1)(\lambda_{\bar{1},3} + \lambda_{2,3} )(\lambda_{1,1} - \lambda_{1,2} - 1)  }} \nn\\
&\qquad\times \sqrt{\frac{ (\lambda_{\bar{1},4} +\lambda_{2,3} + 1 )(\lambda_{\bar{2},4} +
\lambda_{2,3} ) }{ (\lambda_{1,2}- \lambda_{2,3} )(\lambda_{1,2} - \lambda_{2,3} - 1)
}},\nn
\end{align}
\begin{align}
N^{3~2~1}_{3~1~1} &= N^1_1 N^2_1 N^3_3 \sqrt{(\lambda_{1,1} - \lambda_{1,2} - 1)^{-1}(\lambda_{1,1} - \lambda_{1,2})^{-1}} \sqrt{(\lambda_{1,2} + \lambda_{\bar{1},3} + 3)^{-1}(\lambda_{1,2} + \lambda_{\bar{1},3} + 2)^{-1}}\nn \\
&= \sqrt{  \frac{ (\lambda_{1,1} - \lambda_{2,2} + 1) (\lambda_{\bar{1},3}+\lambda_{1,3} + 1 )^2(\lambda_{1,3} - \lambda_{1,1} )(\lambda_{1,3} + \lambda_{\bar{1},3} + 2)(\lambda_{2,3} + \lambda_{\bar{1},3} )  }
{ \left(\lambda_{2,3}-\lambda_{1,3} - 1 \right)(\lambda_{1,2} + \lambda_{\bar{1},3} + 2)(\lambda_{1,4} + \lambda_{\bar{1},3} +
1)(\lambda_{2,2} + \lambda_{\bar{1},3} + 1)(\lambda_{2,4} + \lambda_{\bar{1},3} )  }} \nn\\
&\qquad\times \sqrt{\frac{ (\lambda_{2,3} + \lambda_{\bar{1},3} + 1)(\lambda_{\bar{1},3} -
\lambda_{\bar{1},4})(\lambda_{\bar{2},4} - \lambda_{\bar{1},3} - 1)  }{ (\lambda_{1,1} -
\lambda_{1,2} - 1)(\lambda_{1,2} + \lambda_{\bar{1},3} +
3)(\lambda_{1,2} + \lambda_{\bar{1},3} + 2)  }},  \nn
\end{align}
\begin{align}
N^{3~2~1}_{1~2~1} &= N^1_1 N^2_2 N^3_1 \sqrt{(\lambda_{1,1} - \lambda_{2,2})^{-1}(\lambda_{1,1} - \lambda_{2,2} + 1)^{-1}} \sqrt{(\lambda_{2,2} + \lambda_{\bar{1},3} + 3)^{-1}(\lambda_{2,2} + \lambda_{\bar{1},3} + 2)^{-1}} \nn \\
&= \sqrt{ \frac{(\lambda_{1,2} - \lambda_{1,1} ) (\lambda_{\bar{1},3}+\lambda_{2,3}
)(\lambda_{1,1} - \lambda_{2,3} + 1 )(\lambda_{2,3} - \lambda_{1,3} - 2)  }
{ \left(\lambda_{1,3}-\lambda_{2,3} + 1 \right) (\lambda_{2,2} - \lambda_{1,3} - 1)(\lambda_{2,4} - \lambda_{1,3} - 2)(\lambda_{\bar{1},3} + \lambda_{1,3} + 2) }} \nn\\
&\qquad\times \sqrt{\frac{(\lambda_{2,3} - \lambda_{1,3} - 1)(\lambda_{\bar{1},4} +
\lambda_{1,3} + 2)(\lambda_{\bar{2},4} + \lambda_{1,3} + 1) }{(\lambda_{\bar{1},3} +
\lambda_{1,3} + 1) (\lambda_{1,1} - \lambda_{2,2})(\lambda_{2,2} + \lambda_{\bar{1},3} +
3)(\lambda_{2,2} + \lambda_{\bar{1},3} + 2)  } }, \nn
\end{align}
\begin{align}
N^{3~2~1}_{2~2~1} &= -N^1_1 N^2_2 N^3_2 \sqrt{(\lambda_{1,1} - \lambda_{2,2})^{-1}(\lambda_{1,1} - \lambda_{2,2} + 1)^{-1}} \nn \\
&= -\sqrt{\frac{ (\lambda_{1,2} - \lambda_{1,1} )(\lambda_{1,1} - \lambda_{2,3} + 1 )(\lambda_{1,3} - \lambda_{2,3} )(\lambda_{1,3} - \lambda_{2,3} + 1)(\lambda_{\bar{1},4} + \lambda_{2,3} + 1 )(\lambda_{\bar{2},4} + \lambda_{2,3} )   }{\left(\lambda_{1,3}-\lambda_{2,3} + 1 \right)(\lambda_{1,2} - \lambda_{2,3} + 1)(\lambda_{1,4} - \lambda_{2,3} )(\lambda_{\bar{1},3} + \lambda_{2,3} + 1) (\lambda_{1,1} - \lambda_{2,2})   }} \nn
\end{align}
and
\begin{align}
N^{3~2~1}_{3~2~1} &= -N^1_1 N^2_2 N^3_3 \sqrt{(\lambda_{1,1} - \lambda_{2,2})^{-1}(\lambda_{1,1} - \lambda_{2,2} + 1)^{-1}} \sqrt{(\lambda_{2,2} + \lambda_{\bar{1},3} + 2)^{-1}(\lambda_{2,2} + \lambda_{\bar{1},3} + 1)^{-1}}\nn \\
&= -\sqrt{\frac{ (\lambda_{1,2} - \lambda_{1,1} )(\lambda_{1,3} + \lambda_{\bar{1},3} + 1)(\lambda_{1,3} + \lambda_{\bar{1},3} + 2)(\lambda_{2,3} + \lambda_{\bar{1},3} )(\lambda_{2,3} + \lambda_{\bar{1},3} + 1)        }{  (\lambda_{1,2} + \lambda_{\bar{1},3} + 2)(\lambda_{1,4} + \lambda_{\bar{1},3} + 1)(\lambda_{2,2} + \lambda_{\bar{1},3} + 1)(\lambda_{2,4} + \lambda_{\bar{1},3} )               }  } \nn\\
&\qquad\times \sqrt{\frac{ (\lambda_{\bar{1},3}+\lambda_{2,3}  )(\lambda_{1,1} - \lambda_{2,3} + 1 )(\lambda_{\bar{1},4} - \lambda_{\bar{1},3})(\lambda_{\bar{2},4} - \lambda_{\bar{1},3} - 1)   }
{ \left(\lambda_{1,3}-\lambda_{2,3} + 1 \right)(\lambda_{1,1} - \lambda_{2,2})(\lambda_{2,2} + \lambda_{\bar{1},3} + 2)(\lambda_{2,2} + \lambda_{\bar{1},3} + 1)       } }. \nn
\end{align}


\section{Concluding remarks}

In this article, we have presented matrix element formulae for type 1 unitary irreducible 
representations of the Lie superalgebra $gl(m|n)$. We make use of
classification results originally presented in the work of Gould and Zhang
\cite{GouZha1990,ZhaGou1990} (summarised in Section \ref{class}) and also rely on the
branching rules presented in Section \ref{branch}. 

Regarding the branching rules, some readers already familiar with similar results on the elementary generators 
from the works of Palev \cite{Palev1987,Palev1989} and Stoilova and Van der Jeugt
\cite{StoiVan2010} may find our branching rules appear too simplistic at first glance. 
One observation is that many of the representations may have a highest weight with
non-trivial component $\omega\delta$ (in the sense of Theorem \ref{secsum}, see Section 
\ref{repcat} for details). Since $\omega\delta$ is the highest weight of a one-dimensional
representation, it will have no effect on the form of the branching rule. In otherwords,
to determine the branching rule of a given irreducible unitary type 1 $gl(m|n)$
representation with highest weight $\Lambda = \Lambda_0 + \gamma \varepsilon + \omega
\delta$, it may be easier to do so by first finding the branching rule of the shifted
highest weight $\Lambda_0 + \gamma \varepsilon$. Indeed, a highest weight of this form will
either be a covariant tensor representation (the branching rule of which is covered in \cite{StoiVan2010}), 
and if not, it will be essentially typical (in fact typical and non-tensorial), and hence the
branching rules in \cite{Palev1989,PaStVa1994} are relevant. Importantly, the branching
rules of these articles coincide with our branching rules of Theorem
\ref{mainbranchingrule}.

The general procedure to find matrices of generators of $gl(m|n+1)$
(including non-elementary ones) corresponding to a type 1 unitary irreducible highest
weight representation is:
\begin{itemize}
\item[1.] Determine the branching rules all the way down the subalgebra chain
(\ref{flag}), using Theorem \ref{mainbranchingrule};
\item[2.]
Express every basis vector as a GT pattern of the form (\ref{genGT});
\item[3.] 
Determine the matrix elements using the formulae presented in Section \ref{resultsum}.
\end{itemize}

From the duality discussed in Section \ref{class}, the dual of a type 1 unitary
irreducible representation 
that is tensorial will be an irreducible type 2 unitary representation that is also tensorial, but in a 
different sense. In this case, the type 2 unitary representations which are tensorial are those
that occur in the tensor product of a number of copies of the {\em dual vector
representation}. Matrix element formulae and related concepts associated with the type 2
unitary irreducible representations will be discussed in detail in a forthcoming article.


%
%
%
%
%
%
%
%

\section*{Appendix A: Branching rule for Kac modules}

Let $\hat{L}=gl(m|n+1)$, with $\mathbb{Z}$-gradation
$$
\hat{L} = \hat{L}_-\oplus \hat{L}_0\oplus \hat{L}_+,\ \ \hat{L}_0 = gl(m)\oplus gl(n+1)
$$
and set $L = gl(m|n)\subset \hat{L},$ with $\mathbb{Z}$-gradation
$$
L = L_-\oplus L_0\oplus L_+,\ \ L_0 = gl(m)\oplus gl(n)\oplus gl(1).
$$

Now given a finite dimensional $\hat{L}$-module $\hat{V}$ we have the $q$-character
$$
\mbox{ch}_q\hat{V} = \sum_\nu m_\nu q^\nu
$$
where the sum is over the distinct weights $\nu$ in $\hat{V}$ each occuring with
multiplicity $m_\nu$.

Of particular interest here is the case of a Kac-Module
$$
\hat{K}(\hat{\Lambda}) = U(\hat{L})\otimes_{\hat{L}_+} \hat{V}_0(\hat{\Lambda})
$$
with $\hat{V}_0(\hat{\Lambda}),$ $\hat{\Lambda}=\hat{\Lambda}^{(0)} + \hat{\Lambda}^{(1)}
\equiv (\hat{\Lambda}^{(0)} | \hat{\Lambda}^{(1)}),$ a finite dimensional irreducible
$\hat{L}_0$-module with highest weight $\hat{\Lambda}.$ Clearly in this case we have
\begin{equation}
\mbox{ch}_q\hat{K}(\hat{\Lambda}) = \hat{D}^q_1\mbox{ch}_q\hat{V}_0(\hat{\Lambda})
\label{star1}
\end{equation}
where ch$_q\hat{V}_0(\hat{\Lambda})$ is the usual $q$-character of the irreducible
$\hat{L}_0$-module $\hat{U}_0(\hat{\Lambda})$ and $\hat{D}^q_1$ is the odd ``denominator''
function
$$
\hat{D}^q_1 = \prod_{\beta\in\hat{\Phi}^+_1}(1+q^{-\beta})
$$
with $\hat{\Phi}^+_1$ the set of odd positive roots of $\hat{L}$.

We have a partition
$$
\hat{\Phi}^+_1 = \Phi^+_1\cup \Delta^+_1
$$
with $\Phi^+_1$ the odd positive roots of $gl(m|n)$ and 
$$
\Delta^+_1 = \{\varepsilon_i - \delta_{m+n+1}\ |\ 1\leq i\leq m \}.
$$
With this notation we may write
$$
\hat{D}^q_1 = D^q_1E^q_1
$$
with $D^q_1$ the corresponding denominator polynomial for $L=gl(m|n)$ and
$$
E^q_1 = \prod_{\beta\in\Delta^+_1}(1+q^{-\beta}).
$$

Now using the usual Gelfand-Tsetlin $\hat{L}_0\downarrow L_0$ branching rules we have the
following decomposition into irreducible  $L_0$-modules:
$$
\hat{V}_0(\hat{\Lambda}^{(0)}|\hat{\Lambda}^{(1)}) = \bigoplus_{\Lambda^{(1)}}\hat{V}_0(\hat{\Lambda}^{(0)}|{\Lambda}^{(1)})  
$$
where the sum is over all $L_0$-highest weights $\Lambda^{(1)}$ subject to the usual
betweenness conditions
$$
\hat{\Lambda}^{(1)}_\mu \geq \Lambda^{(1)}_\mu \geq \hat{\Lambda}^{(1)}_{\mu +1},\ \ 1\leq
\mu\leq n.
$$
This immediately gives
\begin{align}
\mbox{ch}_q\hat{K}(\hat{\Lambda})  &= \hat{D}^q_1\sum_{\Lambda^{(1)}} \mbox{ch}_q
V_0(\hat{\Lambda}^{(0)}|\Lambda^{(1)}) \nonumber\\
&= D^q_1\sum_{\Lambda^{(1)}} E^q_1\mbox{ch}_q V_0(\hat{\Lambda}^{(0)}|\Lambda^{(1)}).
\label{star2}
\end{align}

Now observe that
$$
E^q_1\mbox{ch}_q V(\hat{\Lambda}^{(0)}|\Lambda^{(1)})
=\sum_{\Lambda^{(0)}} \mbox{ch}_q V(\Lambda^{(0)}|\Lambda^{(1)})
$$
where the sum is over all $gl(m)$ dominant weights $\Lambda^{(0)}$ such that
$$
\hat{\Lambda}^{(0)}_i\geq \Lambda^{(0)}_i\geq \hat{\Lambda}^{(0)}_i-1.
$$
Substituting into equation (\ref{star2}) we arrive at
$$
\mbox{ch}_q\hat{K}(\hat{\Lambda}) = D^q_1\sum_{\Lambda}\mbox{ch}_qV_0(\Lambda)
$$
where the sum is over all $L_0$ highest weights $\Lambda=(\Lambda^{(0)}|\Lambda^{(1)})$
subject to the conditions
\begin{align}
\hat{\Lambda}^{(0)}_i\geq \Lambda^{(0)}_i\geq \hat{\Lambda}^{(0)}_i-1, & \ \ 1\leq i\leq
m, \label{star31}\\
\hat{\Lambda}^{(1)}_\mu\geq \Lambda^{(1)}_\mu\geq \hat{\Lambda}^{(1)}_\mu, & \ \ 1\leq \mu\leq
n. \label{star32}
\end{align}
These are precisely the branching conditions presented in our previous paper \cite{GIW1}.

In terms of Kac-modules, let
$$
K(\Lambda) = U(L_-)\otimes_{L_+} V_0(\Lambda)
$$
be a corresponding Kac-module for $L=gl(m|n).$ Then the above shows that
\begin{equation}
\mbox{ch}_q\hat{K}(\hat{\Lambda}) = \sum_\Lambda\mbox{ch}_qK(\Lambda),
\label{star4}
\end{equation}
i.e. the branching condition of our previous paper \cite{GIW1} coincides precisely with
the decomposition of Kac-modules in the above sense.

In the case of an essentially typical irreducible $\hat{L}$-module
$\hat{V}(\hat{\Lambda})$ we have
\begin{align*}
\hat{V}(\hat{\Lambda}) &= \hat{K}(\hat{\Lambda})\\
&= \bigoplus_{\Lambda} V(\Lambda)
\end{align*}
where the sum is over all $\Lambda$ subject to the betweenness conditions
(\ref{star31}),(\ref{star32}) of our previous paper.

{\bf Remark:} 
We emphasize that equation (\ref{star4}) holds for general dominant
$\hat{\Lambda}$.


\section*{Appendix B: Phase convention}

In this appendix we derive the phase of the matrix elements of the generators $E_{p,p+2}$ and then extend this result to matrix elements of all generators $E_{p,p+q}$.

The simple generators $E_{p,p+1}$ acting on a GT pattern $|\Lambda \rangle$
($\Lambda$ the highest weight of a type 1 unitary representation for $gl(m|n+1)$) will produce
\begin{align}
E_{p,p+1} |\Lambda \rangle = \sum^p_{a=1} N^p_a |\Lambda + \varepsilon_{a,p} \rangle \nn
\end{align}
where $|\Lambda + \varepsilon_{a,p} \rangle$ is the GT pattern $|\Lambda\rangle$ but with
the $a$th label of the $p$th row shifted by $+1$.
Consequently, non-zero matrix elements of the simple generators will be of the form
\begin{align}
\langle \Lambda + \varepsilon_{a,p} | E_{p,p+1} |\Lambda \rangle  = + N^p_a [\Lambda] \label{BasicME}
\end{align}
where we have set $N^p_a$ to be positive by the Condon-Shortly convention. Non-zero matrix elements of non-simple generators $E_{p,p+2}$ are given by
\begin{align}
N^{p~p+1}_{a~b} = \langle \Lambda + \varepsilon_{a,p} + \varepsilon_{b,p+1} | E_{p,p+2} | \Lambda \rangle &= \langle \Lambda + \varepsilon_{a,p} + \varepsilon_{b,p+1} | [E_{p,p+1} , E_{p+1,p+2}] | \Lambda \rangle \nn\\ 
 &= \langle \Lambda + \varepsilon_{a,p} + \varepsilon_{b,p+1} | E_{p,p+1} | \Lambda + \varepsilon_{b,p+1} \rangle \langle \Lambda + \varepsilon_{b,p+1} | E_{p+1,p+2} | \Lambda \rangle \nn\\
 &~-  \langle \Lambda + \varepsilon_{a,p} + \varepsilon_{b,p+1} | E_{p+1,p+2} | \Lambda + \varepsilon_{a,p} \rangle \langle \Lambda + \varepsilon_{a,p} | E_{p,p+1}  | \Lambda \rangle. \nn
\end{align}
Using (\ref{BasicME}) the above equation can be written as 
\begin{align}
N^{p~p+1}_{a~b} = N^p_a [\Lambda + \varepsilon_{b,p+1} ] N^{p+1}_b [\Lambda] -  N^{p+1}_b [\Lambda + \varepsilon_{a,p}] N^p_a [\Lambda] \nn
\end{align}
where all of the matrix elements on the RHS are positive due to the Baird-Beidenharn convention.
From our previous results \cite{GIW1}
\begin{align}
\bar{\delta}_a &= (-1)^{|I'|} \prod_{b\in I',b\neq a}\left( \bar{\alpha}_b-\bar{\alpha}_a +(-1)^{(b)}
\right)^{-1} \prod_{c\in \tilde{I}'}\left( \bar{\alpha}_a - \bar{\beta}_c \right), a\in I, \nn
\end{align}
and
\begin{align}
\bar{c}_a = \prod_{k\in \tilde{I}',k\neq a} \left(\bar{\beta}_a - \bar{\beta}_k\right)^{-1}\prod_{r\in
I'} \left(\bar{\beta}_a - \bar{\alpha}_r - (-1)^{(r)}\right),\ \ a\in \tilde{I}'.
\nn
\end{align}
So for odd $a$ and odd $b$
\begin{align*}
N^{p~p+1}_{a~b} [\Lambda] &= N^p_a [\Lambda + \varepsilon_{b,p+1} ] N^{p+1}_b [\Lambda] -  N^{p+1}_b [\Lambda + \varepsilon_{a,p}] N^p_a [\Lambda] \nn\\
&= (\bar{\delta}_{a,p} \bar{c}_{a,p})^{1/2} [\Lambda + \varepsilon_{b,p+1} ] N^{p+1}_b [\Lambda] -  (\bar{\delta}_{b,p+1} \bar{c}_{b,p+1})^{1/2} [\Lambda + \varepsilon_{a,p}] N^p_a [\Lambda] \nn\\
&= (\bar{\delta}_{a,p})^{1/2} [\Lambda + \varepsilon_{b,p+1} ] (\bar{c}_{a,p})^{1/2} [\Lambda]  N^{p+1}_b [\Lambda] -  (\bar{c}_{b,p+1})^{1/2} [\Lambda + \varepsilon_{a,p}] (\bar{\delta}_{b,p+1})^{1/2} [\Lambda] N^p_a [\Lambda] \nn\\
&= \frac{ \sqrt{\bar{\alpha}_a - \bar{\beta}_b - 1} }{ \sqrt{\bar{\alpha}_a - \bar{\beta}_b} } (\bar{\delta}_{a,p})^{1/2} (\bar{c}_{a,p})^{1/2} [\Lambda]  N^{p+1}_b [\Lambda] -  \frac{\sqrt{\bar{\beta}_b - \bar{\alpha}_a}}{\sqrt{\bar{\beta}_b - \bar{\alpha}_a + 1} }  (\bar{c}_{b,p+1})^{1/2} (\bar{\delta}_{b,p+1})^{1/2} [\Lambda] N^p_a [\Lambda] \nn\\
&= \left( \sqrt{\frac{ \bar{\alpha}_a - \bar{\beta}_b - 1 }{ \bar{\alpha}_a - \bar{\beta}_b }} -  \sqrt{\frac{\bar{\alpha}_a - \bar{\beta}_b }{\bar{\alpha}_a - \bar{\beta}_b - 1} } \right)  N^p_a N^{p+1}_b  [\Lambda] \nn\\
&= -( \bar{\alpha}_a - \bar{\beta}_b - 1)^{-1/2} (\bar{\alpha}_a - \bar{\beta}_b)^{-1/2} N^p_a N^{p+1}_b  [\Lambda] \nn\\
&= ( \bar{\beta}_b - \bar{\alpha}_a + 1)^{-1/2} (\bar{\beta}_b - \bar{\alpha}_a )^{-1/2} N^p_a N^{p+1}_b  [\Lambda] \nn
\end{align*}
which matches equation (\ref{NonSimpleRaising2}) for $l = p-1$. Similarly, for the cases
corresponding to the other three parity combinations of $a$ and $b$, we obtain the same result.

We observe that the sign of $N^{p~p+1}_{a~b}$ is directly given by the sign of
$\bar{\beta}_b - \bar{\alpha}_a$. In fact, the sign of $N[u_p,u_{p-1},\ldots,u_l]$ is given by
the multiplied signs of the $\bar{\rho}$ terms in (\ref{NonSimpleRaising}). Note that the
above derivation implies that the sign of $(\rho_{u_s,u_{s-1}})^{1/2}$ is given by taking
the square root of $( \bar{\beta}_{u_s} - \bar{\alpha}_{u_{s-1}} + 1)$ and $(
\bar{\beta}_{u_s} - \bar{\alpha}_{u_{s-1}})$ \textit{individually}.

We therefore have
\begin{align}
S(N[u_p,u_{p-1},\ldots,u_l]) &= \prod^p_{s=l+1} S(\bar{\rho}_{u_s,u_s-1}) \nn\\
&= \prod^p_{s=l+1} S(\bar{\beta}_{u_s} - \bar{\alpha}_{u_s-1}). \label{ProductOfPhases}
\end{align}
For $(u_s) = 0$, $(u_{s-1})=0$,$u_s \neq u_{s-1}$
\begin{align}
S(\bar{\beta}_{u_s} - \bar{\alpha}_{u_{s-1}}) &= S(\Lambda_{u_{s-1}} - \tilde{\Lambda}_{u_s}  + u_s - u_{s-1}) \nn\\
&= S(u_s-u_{s-1}). \nn
\end{align}
For $(u_s) = 1$, $(u_{s-1})=1$
\begin{align}
S(\bar{\beta}_{u_s} - \bar{\alpha}_{u_{s-1}}) &= S(\tilde{\Lambda}_{u_s} - \Lambda_{u_{s-1}} + u_{s-1} - u_s) \nn\\
&= S(u_{s-1}-u_s). \nn
\end{align}
For $(u_s) = 1$, $(u_{s-1})=0$
\begin{align}
S(\bar{\beta}_{u_s} - \bar{\alpha}_{u_{s-1}}) &= S(\bar{\beta}_{u_s} - \bar{\beta}_{u_s-1} - 1) \nn\\
&= S(~(\Lambda + \rho, \epsilon_{u_{s-1}} - \delta_{u_s} )~). \nn
\end{align}
For $\Lambda$ typical type 1 unitary we have $(\Lambda + \rho,\epsilon_m - \delta_n) > 0$ which gives
\begin{align}
(\Lambda + \rho, \epsilon_{u_{s-1}} - \delta_{u_s} ) &= (\Lambda + \rho, \epsilon_m -
\delta_n ) + (\Lambda + \rho, \epsilon_{u_{s-1}} - \epsilon_m ) + (\Lambda + \rho,
\delta_n - \delta_{u_s}) \nn\\
&  \geq (\Lambda + \rho,\epsilon_m - \delta_n) > 0. \nn
\end{align}
For $\Lambda$ atypical type 1 unitary there exists an odd index $1 \leq \mu \leq n$ such that $(\Lambda + \rho,\epsilon_m - \delta_\mu) = 0$ and $(\Lambda,\delta_\mu - \delta_n) = 0$. Since the labels $\Lambda_\nu$ for $\mu \leq \nu \leq n$ are all equal, only odd labels $\Lambda_{u_s}$ for $u_s \leq \mu$ may be raised. For this matrix element we necessarily have $u_s \leq \mu$ giving
\begin{align}
(\Lambda + \rho, \epsilon_{u_{s-1}} - \delta_{u_s} ) &= (\Lambda + \rho, \epsilon_m - \delta_\mu ) + (\Lambda + \rho, \epsilon_{u_{s-1}} - \epsilon_m ) + (\Lambda + \rho, \delta_\mu - \delta_{u_s}) \nn\\
&= (\Lambda + \rho, \epsilon_{u_{s-1}} - \epsilon_m ) + (\Lambda + \rho, \delta_\mu -
\delta_{u_s}) \geq 0, \nn
\end{align}
which shows that for this case the matrix element is positive, i.e.
\begin{align}
S(\bar{\beta}_{u_s} - \bar{\alpha}_{u_s-1}) &= 1,~~~~(u_s) = 1,(u_{s-1})=0, \nn
\end{align}
and similarly
\begin{align}
S(\bar{\beta}_{u_s} - \bar{\alpha}_{u_s-1}) &= -1,~~~~(u_s) = 0,(u_{s-1})=1. \nn
\end{align}
Combining the above four cases gives
\begin{align}
S(\bar{\beta}_{u_s} - \bar{\alpha}_{u_{s-1}}) =  (-1)^{(u_s-1)(u_s)} S(u_s - u_{s-1}) \nn
\end{align}
where, as usual, odd indices are considered greater than even indices. Finally, from equation (\ref{ProductOfPhases}) we have the result
\begin{align}
S(N[u_p,u_{p-1},\ldots,u_l]) &= \prod^p_{s=l+1} (-1)^{(u_{s-1})(u_s)} S(u_s - u_{s-1}). \nn
\end{align}


\end{document}